\documentstyle[aps,preprint]{revtex}
\begin{document}
\draft
\tighten
\preprint{\vbox{
                \hfill TIT/HEP-441/NP 
 }}
\title{The $F/D$ ratio and meson-baryon couplings from QCD sum rules-II}
\author{Hungchong Kim$^1$ \footnote{E-mail : 
hckim@th.phys.titech.ac.jp, JSPS fellow},
Takumi Doi and
Makoto Oka }
\address{Department of Physics, Tokyo Institute of Technology, Tokyo 
152-8551, 
Japan }
\author{Su Houng Lee}
\address{Department of Physics and 
Institute of Physics and 
Applied Physics,
 Yonsei University, Seoul 120-749, Korea}

\maketitle
\begin{abstract}
Using QCD sum rules, we 
compute the diagonal meson-baryon couplings, 
$\pi NN$, $\eta NN$, $\pi \Xi\Xi$, $\eta \Xi\Xi$,
$\pi \Sigma \Sigma$ and $\eta \Sigma \Sigma$, from the
baryon-baryon correlation function with a meson,
$i\int d^4x~ e^{iq\cdot x} \langle 0| T J_B(x) {\bar J}_B(0)|M(p)\rangle$.
The calculations are performed to leading order in $p_\mu$ by
considering the two separate Dirac 
structures, $i \gamma_5 \gamma_\mu p^\mu$ and
$\gamma_5 \sigma_{\mu \nu} {q^\mu p^\nu}$ separately.
We first improve the previous sum rule calculations on 
these Dirac structures
for the $\pi NN$ coupling by including three-particle pion wave 
functions of twist 4 and then extend the formalism to calculate 
the other couplings,
$\eta NN$, $\pi \Xi\Xi$, $\eta \Xi\Xi$,
$\pi \Sigma \Sigma$ and $\eta \Sigma \Sigma$.
In the SU(3) symmetric limit, we identify the terms
responsible for the $F/D$ ratio in the OPE by matching the
obtained couplings with their SU(3) relations.  Depending on the
Dirac structure considered, we find different identifications
for the $F/D$ ratio.  
The couplings including the SU(3) breaking effects are
also discussed within our approach.

\end{abstract}

\section{INTRODUCTION}
\label{sec:intro}

The QCD sum rule~\cite{SVZ} is often used to determine  
hadronic parameters from QCD.  
In this framework, an interpolating field appropriate for
the hadron of concern is introduced using quark and gluon
fields and used to construct an appropriate correlation function.
The correlation function is calculated on the one hand by
the operator product expansion (OPE) at the deep 
Euclidean region of the correlator momentum
$q^2\rightarrow -\infty$ using QCD degrees
of freedom.  On the other hand, its phenomenological
form is constructed using hadronic degrees of freedom.
In most cases of practical QCD sum rule calculations,  the 
phenomenological form is analytic
in the complex $q^2$ plane except along the
positive real axis.   Through a dispersion relation,
the nonanalytic structure along the
positive real axis is matched  to the QCD representation of
the correlator at $q^2\rightarrow -\infty$ and
the hadron parameter of concern is extracted in terms
of QCD parameters. 

The QCD sum rule framework has been widely used to
calculate various hadronic properties~\cite{qsr}.  
Among various applications, 
determining meson-baryon couplings is of particular interest
because meson-baryon couplings are important ingredients for analyzing
baryon-baryon interactions.  Their values determined from QCD may
provide important constraints in constructing baryon-baryon 
potentials~\cite{bonn}.
One main feature of meson-baryon couplings is
SU(3) symmetry as it provides a systematic classification of 
the couplings~\cite{swart}  in terms of the 
two parameters, the $\pi NN$ coupling and the $F/D$ ratio.
This systematic classification of the couplings is a basis
for making realistic potential models for 
hyperon-baryon interactions~\cite{rijken1,rijken2}.  
In this approach, however, implementing the SU(3) breaking 
in the couplings is somewhat limited because the models used
rely solely on hadronic degrees of freedom and thus the way of
introducing
the SU(3) breaking terms in the model may not be unique.
Moreover, the baryon-baryon scattering data used in the fitting
processes are not precise enough to pick out a specific
mesonic channel.  Therefore, 
it would be useful to constrain each model of meson-baryon coupling
directly from other non-perturbative methods of QCD, such as QCD sum rules.

Recently, the two-point correlation function of 
the nucleon interpolating fields with an external pion field
\begin{eqnarray}
\Pi (q, p) = i \int d^4 x e^{i q \cdot x} \langle 0 | T[J_N (x) 
{\bar J}_N (0)]| \pi (p) \rangle \ .
\label{two}
\end{eqnarray}
has been extensively used to calculate the pion-nucleon coupling within
the conventional QCD sum rule 
method~\cite{hung1,hung2,hung5,hung6,hat,krippa}.
Another approach relying on the three-point function~\cite{yazaki}
gives results that may contain non-negligible contributions from
the higher resonances $\pi(1300)$ and $\pi(1800)$~\cite{maltman}.
Moreover, using the two-point correlation function, the sum rule can
be easily extended to other meson-baryon couplings and the SU(3)
limit can be easily taken to identify the $F/D$ ratio.
Among various developments using Eq.~(\ref{two}), one interesting attempt 
is to calculate the coupling beyond the
chiral limit~\cite{hung5} by considering the Dirac structure
$i\gamma_5$ at the order $p^2=m_\pi^2$. 
As a consistent chiral counting, linear terms in the quark mass $m_q$
have been included in the OPE side. 
The pion-nucleon coupling obtained by combining this sum rule with
the nucleon chiral-odd sum rule seems to be quite satisfactory.

This sum rule beyond the chiral limit has been recently applied to 
other meson-baryon
couplings such as $\eta NN$, $\pi \Xi\Xi$, $\eta \Xi\Xi$, $\pi \Sigma \Sigma$
and $\eta \Sigma \Sigma$~\cite{hung6}. 
The OPE of each sum rule is found to satisfy
the SU(3) relations for the couplings proposed in Ref.~\cite{swart},
which enables us to identify the terms responsible for the $F/D$ ratio.
This nontrivial observation for the $F/D$ ratio, which is 
a natural consequence 
of using the SU(3) symmetric interpolating fields,
was possible because the sum rules are constructed beyond the
chiral limit.   The sum rules, if constructed in the
soft-meson limit for example, provide the OPE trivially
satisfying the SU(3) relations with $F/D=0$.  Therefore, going
beyond the chiral limit especially in the $i\gamma_5$ sum rules  
is important for obtaining 
nontrivial value of the $F/D$ ratio.
From this sum rule, the ratio obtained is $F/D \sim 0.2$
substantially smaller than what it has been known from SU(6)
consideration. Furthermore,
meson-baryon couplings after taking into account the SU(3) breaking in 
the OPE as well as in the phenomenological part
undergo huge changes from their SU(3) symmetric values. 
This finding is not consistent with Nijmegen 
potentials~\cite{rijken1,rijken2} or with common assumptions in studying 
hypernuclei.

On the other hand, similar sum rule calculations can be performed 
for the couplings using Dirac structures other than the $i\gamma_5$
structure.  As discussed in Ref.~\cite{hung2},
the correlation function Eq.~(\ref{two}) contains the other
Dirac structures, $i\gamma_5 {\hat p}$ (${\hat p} \equiv \gamma_\mu p^\mu $ )
and $\gamma_5 \sigma_{\mu \nu} q^\mu p^\nu$,
which can also be used to construct 
sum rules for the $\pi NN$ coupling beyond the soft-pion limit.  
Of course, it is straightforward to extend the
$i\gamma_5 {\hat p}$ and $\gamma_5 \sigma_{\mu \nu} q^\mu p^\nu$
sum rules and calculate the couplings
$\eta NN$, $\pi \Xi\Xi$, $\eta \Xi\Xi$, 
$\pi \Sigma \Sigma$and $\eta \Sigma \Sigma$.   In the
SU(3) limit, the OPE should satisfy the SU(3) relations for the
couplings, which will allow us to identify the OPE
terms responsible for the
$F/D$ ratio.
This is our primary purpose of this work.
In particular, we will see if the identifications made from
these Dirac structures are consistent
with the previous identification made in Ref.~\cite{hung6},
or if the Dirac structure dependence of the sum rule results~\cite{hung2}
still persists in these identifications. 
Once the sum rules are constructed, it will be straightforward
to introduce the
SU(3) breaking within this framework and see how the SU(3) breaking
affects the couplings.

As presented in Ref.~\cite{hung1}, the 
$\gamma_5 \sigma_{\mu \nu} q^\mu p^\nu$ Dirac structure
has nice features in calculating the $\pi NN$ coupling.
To be specific, this sum rule provides a coupling
independent of the models employed to construct its
phenomenological side and the result is rather stable against the
variation of the continuum
parameter. Indeed, this structure has been further 
applied to the study of the
couplings $g_{NK\Lambda}$ and $g_{NK\Sigma}$~\cite{bracco} and
other pion-baryon couplings~\cite{aliev}.
Even though the calculated $\pi NN$ coupling is a bit smaller than 
the empirical value, it may
be useful to investigate this sum rule further and improve the previous
result.  In this work, we will revisit the previous $\pi NN$ calculations 
and improve the OPE calculation in the highest dimensions.
We then apply the framework to other meson-baryon couplings
in the SU(3) sector to identify the OPE responsible for the $F/D$ ratio.

The other Dirac structure $i\gamma_5 {\hat p}$, 
as discussed in Ref.~\cite{hung2}, is found not to be reliable
for calculating the $\pi NN$ coupling as it
contains large contributions from the continuum.
The extracted value is highly sensitive to the continuum threshold.
Nevertheless, we will study this sum rule again
for the completeness first of all.   
We will improve the OPE calculation by including higher dimensional
operators.  Then, by extending the sum rule to the
other couplings,  we will see how the 
identification for the $F/D$ ratio from this Dirac structure
is different from the ones obtained from other Dirac structures.

The paper is organized as follows.  In Section~\ref{sec:pnn},
we will revisit the $i\gamma_5 {\hat p}$ and 
$\gamma_5 \sigma_{\mu\nu} q^\mu p^\nu$ sum rules for $\pi NN$
and refine the OPE calculation.  We then extend this framework
to construct $\eta NN$ sum rules in Section~\ref{sec:enn}
and identify the OPE responsible for the $F/D$ ratio.
In Section~\ref{sec:others}, we construct sum rules
for the couplings, $\pi \Xi \Xi$, $\eta \Xi \Xi$, 
$\pi \Sigma \Sigma$ and $\eta \Sigma \Sigma$.  
We confirm that the OPE of each sum rule satisfies the
SU(3) relations with the identification of the $F/D$ ratio made
in Section~\ref{sec:enn}. In Section~\ref{sec:su3}, we
present our analysis in the SU(3) symmetric limit. 
The analysis beyond the SU(3) limit is given in Section~\ref{sec:bsu3}
but only for the $\gamma_5 \sigma_{\mu\nu} q^\mu p^\nu$ sum rules.

\section{Pion-nucleon coupling determined beyond the soft-pion limit}
\label{sec:pnn}

In this section, we construct QCD sum rules for the 
$\pi^0 pp$ coupling at the first order of the pion momentum
${\cal O}(p_\mu)$, to predict the coupling beyond the
soft-pion limit.  To do this, we use
the two-point correlation function with a pion,
\begin{eqnarray}
\Pi (q, p) = i \int d^4 x e^{i q \cdot x} \langle 0 | T[J_p (x) 
{\bar J}_p (0)]| \pi^0 (p) \rangle \ 
\equiv \int d^4 x e^{i q \cdot x} \Pi (x,p)\ .
\label{ptwo}
\end{eqnarray}
Here $J_p$ is the proton interpolating field suggested by 
Ioffe~\cite{ioffe1},
\begin{eqnarray}
J_p = \epsilon_{abc} [ u_a^T C \gamma_\mu u_b ] \gamma_5 \gamma^\mu d_c\ ,
\end{eqnarray}
where $a,b,c$ are color indices, $T$ denotes the transpose with
respect to the Dirac indices, $C$ the charge 
conjugation. Using this correlator, we construct the sum rules beyond
the soft-pion limit by considering the Dirac structures, 
$i\gamma_5 {\hat p}$ (${\hat p} \equiv \gamma_\mu p^\mu$) and 
$\gamma_5\sigma_{\mu\nu} q^\mu p^\nu$. After taking out the
Dirac structures containing one power of the pion momentum, we take the 
limit $p_\mu \rightarrow 0$ in the rest of the correlator.
These sum rules have been considered in Ref.~\cite{hung1,hung2}
but we revisit them here to improve the OPE calculations.
By considering the sum rules for $\pi^0 pp$  instead
of $\pi^+ pn$,
we can easily extend the formalism to the other 
diagonal meson-baryon couplings, $\eta NN$, $\pi \Xi \Xi$, $\eta \Sigma \Sigma$
and so on.

The QCD side of the correlator Eq.(\ref{ptwo}) is calculated via
the operator product expansion (OPE) at the deep spacelike region
$q^2\rightarrow -\infty$.  In our calculations,
we use the vacuum saturation hypothesis to factor out quark-antiquark
component with a pion (denoted by $D^q_{ab}$ below) from the correlator.
The rest of the correlator is basically time-ordered products of quark fields 
which are normally evaluated by background-field techniques~\cite{nov1}. 
Accordingly, it is straightforward 
to write the correlator in the coordinate space,
\begin{eqnarray}
\Pi(x,p) = -i \epsilon_{abc} \epsilon_{a'b'c'}
\Bigg \{&& \gamma_5 \gamma^\mu D^d_{cc'} \gamma^\nu \gamma_5
{\rm Tr} \left[i S_{aa'}(x)(\gamma_\nu C)^T iS_{bb'}^T(x) 
(C\gamma_\mu)^T \right ] \nonumber \\
&-&
\gamma_5 \gamma^\mu D^d_{cc'} \gamma^\nu \gamma_5
{\rm Tr} \left[i S_{ab'}(x)\gamma_\nu C iS_{ba'}^T(x) 
(C\gamma_\mu)^T \right ]\nonumber \\
&-&
\gamma_5 \gamma^\mu iS_{cc'}(x) \gamma^\nu \gamma_5
{\rm Tr} \left[i S_{ab'}(x)\gamma_\nu C (D^u_{ba'})^T    
(C\gamma_\mu)^T \right ]\nonumber \\
&+&
\gamma_5 \gamma^\mu iS_{cc'}(x) \gamma^\nu \gamma_5
{\rm Tr} \left[i S_{aa'}(x) (\gamma_\nu C)^T (D^u_{bb'})^T    
(C\gamma_\mu)^T \right ]\nonumber \\
&+&
\gamma_5 \gamma^\mu iS_{cc'}(x) \gamma^\nu \gamma_5
{\rm Tr} \left[ D^u_{aa'}(\gamma_\nu C)^T iS_{bb'}^T(x)    
(C\gamma_\mu)^T \right ]\nonumber \\
&-&
\gamma_5 \gamma^\mu iS_{cc'}(x) \gamma^\nu \gamma_5
{\rm Tr} \left[ D^u_{ab'}\gamma_\nu C iS_{ba'}^T(x)    
(C\gamma_\mu)^T \right ]
\Bigg \}\ .
\label{xp}
\end{eqnarray}
The quark propagators $iS(x)$\footnote{
See Ref.~\cite{wilson} for a detailed expression for the quark
propagator.} inside the
traces are the u-quark propagators and the ones outside of
the traces are the d-quark propagators. For the time being,
we postpone discussions on the gluonic contributions which are obtained by
moving a gluon tensor from a
quark propagator into the quark-antiquark component with
a pion (constituting thus the three-particle pion wave functions 
according to the 
nomenclature commonly used in the light-cone QCD sum rules~\cite{bely}). 
Then, it is possible to write the quark-antiquark component with a pion
in terms of three matrix elements involving a pion.  Namely, for $q=
{u~{\rm or}~d}$-quark, we have 
\begin{eqnarray}
(D_{aa'}^q)^{\alpha \beta} &\equiv&
\langle 0 | q^\alpha_a (x) {\bar q}^\beta_{a'} (0) | \pi^0 (p) \rangle\ 
\nonumber \\
& = & {\delta_{a a'} \over 12}
(\gamma^\mu \gamma_5)^{\alpha \beta}
\langle 0 |
{\bar q} (0) \gamma_\mu \gamma_5  q (x) | \pi^0 (p) \rangle\
+ {\delta_{a a'} \over 12 } (i \gamma_5)^{\alpha \beta}
\langle 0 |
{\bar q}(0) i \gamma_5  q (x) | \pi^0 (p) \rangle
\nonumber \\
&& - {\delta_{a a'} \over 24} (\gamma_5 \sigma^{\mu\nu})^{\alpha\beta}
\langle 0 |
{\bar q}(0) \gamma_5 \sigma_{\mu\nu}  q (x) | \pi^0 (p) \rangle\ .
\label{dd}
\end{eqnarray}
The pseudoscalar pion matrix element $\langle 0 | 
{\bar q}(0) i \gamma_5  q (x) | \pi^0 (p) \rangle$ contributes
only to the $i\gamma_5$ structure of the correlator, not participating
in the sum rules of the Dirac structures
$i\gamma_5 {\hat p}$ and 
$\gamma_5\sigma_{\mu\nu} q^\mu p^\nu$.
The rest two components differ by their chirality 
and both contribute to the $i\gamma_5 {\hat p}$ 
and $\gamma_5\sigma_{\mu\nu} q^\mu p^\nu$ sum rules.

At the first order in $p_\mu$, the pseudovector matrix element
for the $u$-quark
up to twist-4 can be calculated as provided in Appendix~\ref{sec:appa},
namely,
\begin{eqnarray}
A_\mu^u (\pi^0) \equiv 
\langle 0| {\bar u}(0) \gamma_\mu \gamma_5 u(x) | \pi^0 (p)\rangle
\sim i f_\pi p_\mu - i{5\over 18} f_\pi \delta^2  
\left ( {1\over 2} x^2 p_\mu  -{1\over 5} x_\mu x\cdot p \right )\ ,
\label{pv1}
\end{eqnarray}
where $f_\pi =93$ MeV and the
twist-4 parameter $\delta^2=0.2$ GeV$^2$ according to Ref.~\cite{nov}. 
The $d$-quark matrix element has the opposite sign from the $u$-quark
element,
\begin{eqnarray}
A_\mu^d (\pi^0) \equiv
\langle 0| {\bar d}(0) \gamma_\mu \gamma_5 d(x) | \pi^0 (p)\rangle
\sim- 
i f_\pi p_\mu + i{5\over 18} f_\pi \delta^2  
\left ( {1\over 2} x^2 p_\mu  -{1\over 5} x_\mu x\cdot p \right )\ ,
\label{pv2}
\end{eqnarray}
because the pion is an isovector particle.
In the local limit (that is, $x_\mu =0$), this equation becomes just 
the PCAC relation.
By taking the divergence of the local operator and applying the soft-pion 
theorem to the resulting matrix element, one can easily derive the well-known 
Gell-Mann$-$Oakes$-$Renner relation,
\begin{eqnarray}
2m_q \langle {\bar q} q \rangle = -m_\pi^2 f^2_\pi \ . 
\end{eqnarray}
This implies that the $m_q$ terms in the OPE are not the same
order as the first order in $p_\mu$. 
In other words, the $m_q$ terms should not be included in the OPE 
when the sum rules are constructed at the first order in $p_\mu$. 
This tricky point does not matter in the
$\pi pp$ or $\eta pp$ sum rules because the $u$ or $d$ quark
masses are small anyway.  However, in cases when
strange quarks are involved, the quark-mass corrections could be compatible
with other OPE terms and therefore this point should be kept in mind:
the sum rules at the first order in $p_\mu$ should not include
quark-mass terms in the OPE.

The other pion matrix element contributing to our sum rules
is the pseudotensor type $B_{\mu\nu}^q (\pi^0) \equiv
\langle 0 | {\bar q}(0) \gamma_5 \sigma_{\mu\nu}  q (x) | \pi^0 (p) \rangle$
with $q=u,d$,
which up to leading order in $p_\mu$ can be written,
\begin{eqnarray}
B_{\mu\nu}^u (\pi^0) \equiv
\langle 0 | {\bar u}(0) \gamma_5 \sigma_{\mu\nu}  u (x) | \pi^0 (p) \rangle
&\rightarrow& -i(p_\mu x_\nu - p_\nu x_\mu) 
{\langle{\bar u} u\rangle \over 6 f_\pi}\nonumber\ ,\\
B_{\mu\nu}^d (\pi^0) \equiv
\langle 0 | {\bar d}(0) \gamma_5 \sigma_{\mu\nu}  d (x) | \pi^0 (p) \rangle
&\rightarrow& + i(p_\mu x_\nu - p_\nu x_\mu) 
{\langle{\bar d} d\rangle \over 6 f_\pi}\ .
\label{pt}
\end{eqnarray}
This will be derived in Appendix~\ref{sec:appb}.
In the sum rules of the Dirac structures $i\gamma_5 {\hat p}$
and $\gamma_5 \sigma_{\mu\nu} q^\mu p^\nu$, 
the matrix elements
$A_\mu^q$ and $B_{\mu\nu}^q$ each multiplied by the
corresponding Dirac matrix according to Eq.~(\ref{dd})
contribute.

\begin{figure}
\caption{ The OPE diagrams that we are considering in this work.
The solid lines denote quark propagators and the wavy lines denote 
gluon lines. The blob in each figure denotes the quark-antiquark component 
interacting
with an external meson (either $\pi$ or $\eta$) field specified by the 
dashed line.
In (a),(b),(c),(d), the blob denotes the pseudovector element $A_\mu^q$ or 
the pseudotensor
element $B_{\mu\nu}^q$, and in (e) and (f), it denotes  
the three-particle wave function 
Eq.~(\ref{three}).  
}
\label{fig1}

\setlength{\textwidth}{6.1in}   
\setlength{\textheight}{9.in}  
\centerline{%
\vbox to 2.4in{\vss
   \hbox to 3.3in{\includegraphics{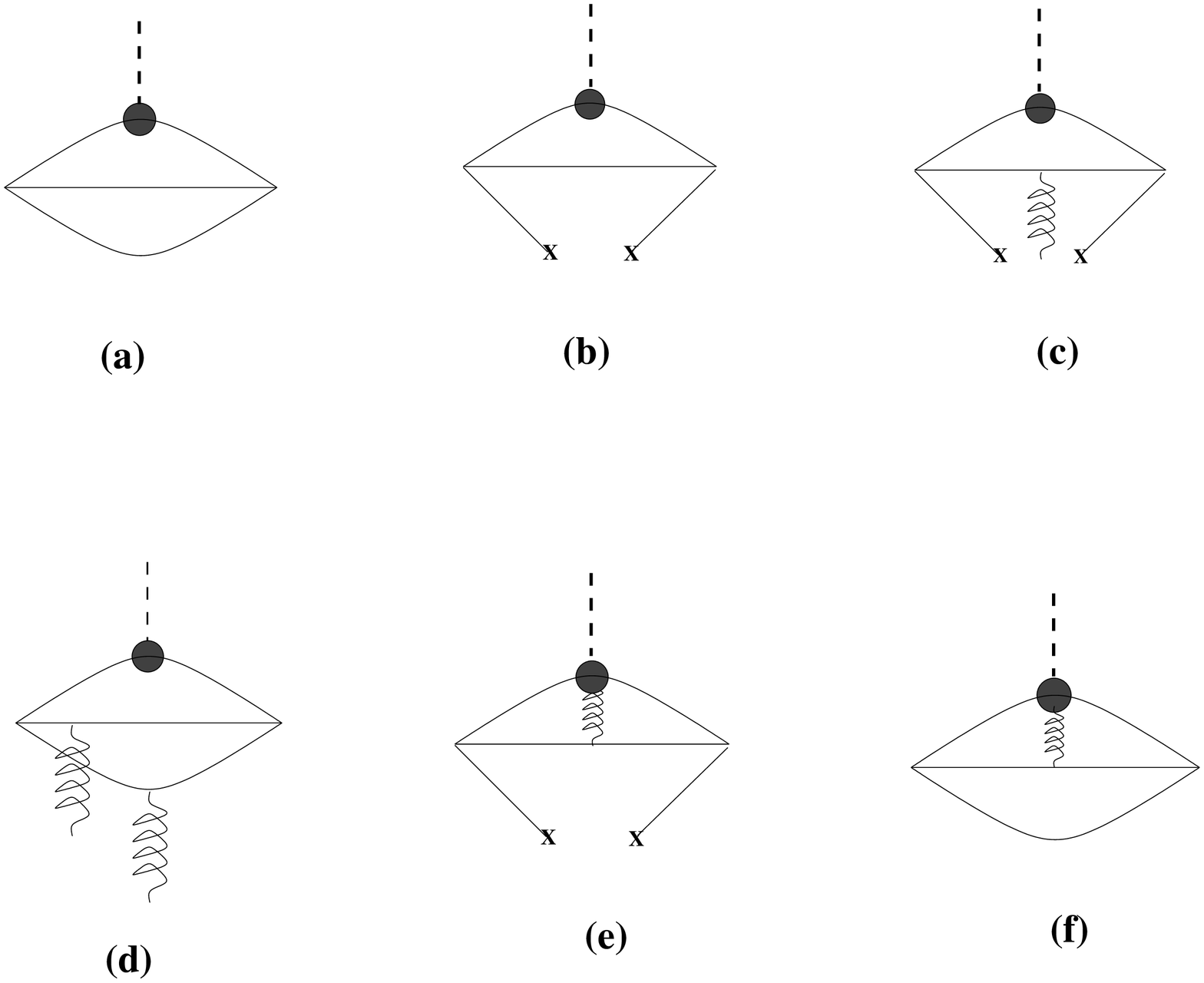}\hss}}
}
\vspace{100pt}
\eject
\end{figure}

The OPE diagrams that we are considering up to dimension 7
are given in figure~\ref{fig1}.  
Each blob in figures~\ref{fig1} [except the figures (e) and (f)]
denotes either $A_\mu^q$ or $B_{\mu\nu}^q$.
For example, if we take the term containing $A_\mu^q$ in Eq.~(\ref{dd}) 
for the pion matrix element,
figure~\ref{fig1} (a) contribute to the $i\gamma_5 {\hat p}$ sum rule.
But in the case of figure~\ref{fig1} (b), even if we take $A_\mu^q$ 
for the blob,
because the chirality is flipped by the disconnected
quark line [namely by the $\langle {\bar q} q \rangle$ condensate], 
this diagram contributes to the 
$\gamma_5 \sigma_{\mu\nu} q^\mu p^\nu$ sum rule.  

Figures~\ref{fig1} (e) (f), where
a gluon from a quark propagator interacts with
the quark-antiquark element with a pion, are new in this work, not 
properly considered 
in our previous calculations~\cite{hung1}.
In these cases, the blob denotes the three-particle pion wave functions
commonly used in the light-cone QCD sum rules~\cite{bely}.
At order $p_\mu$, the only three-particle wave function
contributing to our sum rules is
\begin{eqnarray}
\langle 0| q(x)^\alpha_a i g_s [{\tilde G}^A (0)]^{\sigma \rho} 
{\bar q}(0)^\beta_b | \pi^0(p) \rangle
&\equiv&-{1\over 16} t^A_{ab} (\gamma_\theta)^{\alpha \beta}
(p^\rho g^{\theta \sigma} - p^\sigma g^{\theta \rho}) A^q_G
\nonumber \\
&\sim& \pm t^A_{ab} (\gamma_\theta)^{\alpha \beta} {f_\pi \delta^2 \over 48} 
(p^\rho g^{\theta \sigma} - p^\sigma g^{\theta \rho})\ ,
\label{three}
\end{eqnarray}
where ${\tilde G}^A_{\alpha\beta}$ is the dual of the gluon
strength tensor, 
${\tilde G}^A_{\alpha\beta}={1\over 2} 
\epsilon_{\alpha\beta\sigma\rho} (G^A)^{\sigma\rho}$.
The plus sign is for the $d$-quark and the minus sign is for the $u$-quark.
For its derivation, see Appendix~\ref{sec:appc}.

Other diagrams contributing to our sum rules but not explicitly shown 
in figures~\ref{fig1} are when a gluonic tensor coming from the disconnected
quark line combines with the quark condensate to form the quark-gluon
mixed condensate.  These are basically obtained from figure~\ref{fig1} (b)
by expanding the disconnected quark line in $x_\mu$. 
We have not drawn those diagrams
since they are basically the same kind as figure~\ref{fig1} (c).  
Similarly, the OPE coming from the coordinate expansion of the quark-antiquark
component with a pion, namely the diagram obtained by 
expanding the blob in 
figure~\ref{fig1} (a) or (b), is the same kind as
figure~\ref{fig1} (e) or (f) and therefore is not explicitly shown.

We now collect the OPE contributing to the $i\gamma_5 {\hat p}$ sum rule first
in the coordinate space and take the Fourier transformation afterward.  
Figure~\ref{fig1} (a) contributes to the  
$i\gamma_5 {\hat p}$ sum rule when the pion matrix element $A_\mu^q$
is taken.  It is
\begin{eqnarray}
{4i\over \pi^4}\gamma_5 \left [ -{x\cdot A^d\over x^8} {\hat x}
+{\hat A^u\over x^6} -{x\cdot A^u\over x^8} {\hat x} \right ]
&=& {4i\over \pi^4}\gamma_5 {\hat A^u\over x^6}
\label{pvope1} \\
&\stackrel{\rm F. T.}\longrightarrow &~~
i\gamma_5 {\hat p} {f_\pi \over 2\pi^2} \left [ q^2 ln(-q^2) + 
\delta^2 ln(-q^2) \right ]\label{pvope12}\ ,
\end{eqnarray}
where we have used the isospin relation 
$A_\mu^u (\pi^0) =-A_\mu^d (\pi^0)$ in the first step and then
Eq.~(\ref{pv1}) afterward before the Fourier transformation is taken. 
In the first step, the $d$-quark contribution was canceled by the
corresponding $u$-quark contribution.
It should be noted however that,
in the case of the $\eta NN$ coupling, such a cancellation does
not occur as $\eta$ is an isoscalar.
The contribution from figure~\ref{fig1} (d) is similarly obtained,
\begin{eqnarray}
&\left \langle {\alpha_s \over \pi} {\cal G}^2
\right \rangle &
{i \gamma_5 \over 144 \pi^2}
\left [ A_\alpha^d {1\over x^4} (x^2 \gamma^\alpha - x^\alpha {\hat x})
-A_\alpha^u {1\over x^4} (-4x^2 \gamma^\alpha + x^\alpha {\hat x})
\right ]
\label{pvope2} \\
&\stackrel{\rm F. T.}\longrightarrow &~~
-i\gamma_5 {\hat p}{f_\pi\over 12}
\left \langle {\alpha_s \over \pi} {\cal G}^2
\right \rangle \left [ -{1\over q^2} - {\delta^2 \over 9 q^4}  \right ]\ .
\label{pvfig1a}
\end{eqnarray}
As in Eq.~(\ref{pvope12}), the isospin relation 
$A_\mu^u (\pi^0) =-A_\mu^d (\pi^0)$ has
been used.
Other diagrams contributing to the $i\gamma_5 {\hat p}$ sum rule
are calculated straightforwardly,
\begin{eqnarray}
&{\rm Fig.\ref{fig1}(f)}& \rightarrow 
i\gamma_5 {\hat p} {3\over 4\pi^2} ln(-q^2)
[A^d_G + A^u_G] = 0 \ ,
\label{pvope3}
\\
&{\rm Fig.\ref{fig1}(b)}& \rightarrow i\gamma_5 {\hat p} 
{2\over 9f_\pi} \langle {\bar u} u\rangle^2 {1\over q^2}\ ,
\label{pvope4}
\\ 
&{\rm Fig.\ref{fig1}(c)}&\rightarrow -i\gamma_5 {\hat p}
{m_0^2 \langle {\bar u} u\rangle^2 \over 36 f_\pi} {1\over q^4}\ .
\label{pvope5}
\end{eqnarray}
In obtaining Eqs.~(\ref{pvope4}) and (\ref{pvope5}), the pseudotensor
pion matrix element $B_{\mu\nu}^q$ has been used for
the blobs in the figures.  The chirality change due to taking
$B_{\mu\nu}^q$ is compensated by the disconnected quark line so that
total chirality is the same as Eq.~(\ref{pvfig1a}).
Note that the quark-gluon mixed condensate in the
last equation is parametrized as
$\langle {\bar u} g_s \sigma \cdot {\cal G} u \rangle \equiv
m_0^2 \langle {\bar u} u\rangle$ with $m^2_0 = 0.8$ GeV$^2$~\cite{ovc}. 

The phenomenological side for the $i\gamma_5 {\hat p}$ sum rule
takes the form,
\begin{eqnarray}
-{g_{\pi N} \lambda_N^2 m_N \over (q^2-m_N^2)^2} + \cdot \cdot \cdot\ .
\end{eqnarray}
The ellipsis includes the continuum whose spectral density is parametrized by a
step function, and the nucleon single-pole term.
The single-pole term comes from $N\rightarrow N^*$ transitions
as well as the PS and PV scheme-dependent $N\rightarrow N$~\cite{hung1}.
As the pion momentum is taken out along with its Dirac structure 
$i\gamma_5 {\hat p}$,
the rest correlator contains only the correlator momentum $q^2$.
We thus employ a single-variable dispersion relation in matching
the two sides of the sum rule.

In QCD sum rules with external fields, a double dispersion relation
is sometime used. (See for example Ref.~\cite{bely}.) 
But as suggested in Ref.~\cite{hung4}, sum rules invoking 
the double dispersion relation may contain spurious terms. 
As discussed in Ref~\cite{hung3}, the spurious terms,
at least in a few examples considered in Ref~\cite{hung3}, give
rise to unphysical poles in the spectral density located at the continuum 
threshold.
Indeed, it was demonstrated in Ref.\cite{hung4} that if sum rules are
constructed 
for each power of the external momentum as we did in this work, 
the sum rules coming from the double dispersion relation are identical 
to that coming from the single dispersion relation, provided the spurious 
terms are eliminated.   
Thus, even though there are on-going discussions~\cite{birse},
we believe that the single-variable dispersion relation leads to the  
correct sum rules constructed at the order $p_\mu$.

Taking the Borel
transformation with respect to the correlator momentum $-q^2$, we 
obtain the $i\gamma_5 {\hat p}$ sum rule,
\begin{eqnarray}
&&g_{\pi N}~m_N~ \lambda_N^2 (1+C_{\pi N} M^2)~e^{-m^2_N/M^2}
=  
{f_\pi\over 2\pi^2} \left [ M^6 E_1 (x) + \delta^2 M^4 E_0 (x) \right ]
\nonumber \\
&&+ M^2 \left [ {f_\pi \over 12}
\left \langle {\alpha_s \over \pi} {\cal G}^2 \right \rangle 
+
{2\langle {\bar q} q \rangle^2  \over 9 f_\pi}  \right ]
-
{\delta^2 f_\pi \over 108}\left \langle {\alpha_s \over \pi} {\cal G}^2
\right \rangle
+
{m_0^2\langle {\bar q} q \rangle^2 \over 36 f_\pi}\ .
\label{pvsum} 
\end{eqnarray}
As the isospin symmetry is always imposed, $\langle {\bar q} q\rangle$
denotes either $\langle {\bar u} u\rangle$ or $\langle {\bar d} d\rangle$.
The contribution from the nucleon single pole term whose residue 
is not known  has been denoted by $C_{\pi N}$.
The continuum contribution is denoted by the factor,
$E_n (x\equiv S_0/M^2) =
1-(1+x+\cdot \cdot\cdot +x^n/n!)e^{-x}$ where $S_0$ is
the continuum threshold.
Comparing the corresponding sum rule in the previous 
calculation~\cite{hung2},  we note that the last two terms are new
in this work.
They however belong to the highest dimension and their magnitudes are 
suppressed by the large numerical factors involved.

We now construct the  
$\gamma_5 \sigma^{\mu\nu} q_\mu p_\nu$ sum rule. This differs from
the $i\gamma_5 {\hat p}$ sum rule by its chirality.
In order to sort out the OPE diagrams contributing to
this sum rule, the total chirality  should be counted carefully.
For example, figure~\ref{fig1} (a)
contributes to the sum rule when the pseudotensor matrix element 
$B^q_{\mu\nu}$ is taken
for the pion matrix element. 
Since the other quark propagators do not change
the chirality, the total chirality is given by the
term containing the matrix element $B^q_{\mu\nu}$.
On the other hand,  
figure~\ref{fig1} (b) contributes to the sum rule
when the pseudovector element $A_\mu^q$ is taken for the pion
matrix element because the chirality change due to this
choice is compensated by the disconnected quark line.
Collecting terms contributing to the  
$\gamma_5 \sigma^{\mu\nu} q_\mu p_\nu$ structure, we have 
\begin{eqnarray}
{\rm Fig.1 (a)}& \stackrel{\rm F. T.}\longrightarrow&
{\langle {\bar d} d \rangle \over 12 \pi^2 f_\pi}
\gamma_5 \sigma^{\mu\nu} q_\mu p_\nu ln(-q^2)\ , 
\label{ptope1}\\
{\rm Fig.1 (b)}& \rightarrow& {2i\over 3\pi^2} \langle {\bar d} d \rangle
{A_\alpha^u \over x^4} x_\beta \gamma_5 \sigma^{\alpha \beta} 
\nonumber \\
& \stackrel{\rm F. T.}\longrightarrow&
{4f_\pi\over 3} \langle {\bar d} d \rangle \gamma_5 
\sigma^{\mu\nu} q_\mu p_\nu
\left ( {1\over q^2} - {5\delta^2 \over 9q^4} \right ) \ ,
\label{ptope2} \\
{\rm Fig.1 (c)}& \rightarrow& {im_0^2\over 48 \pi^2} \langle {\bar d} d \rangle
A_\alpha^u \gamma_5 \sigma^{\alpha \beta} {x_\beta \over x^2} 
\nonumber \\
& \stackrel{\rm F. T.}\longrightarrow&
{f_\pi m_0^2 \over 6} \langle {\bar d} d \rangle 
\gamma_5 \sigma^{\mu\nu} q_\mu p_\nu {1\over q^4} \ ,
\label{ptope3} \\
{\rm Fig.1 (d)}& \stackrel{\rm F. T.}\longrightarrow&
-{1\over 216 f_\pi} \langle {\bar d} d \rangle  
\left \langle {\alpha_s \over \pi} {\cal G}^2 \right \rangle 
\gamma_5 \sigma^{\mu\nu} q_\mu p_\nu {1\over q^4}\ ,
\label{ptope4}\\
{\rm Fig.1 (e)}& \stackrel{\rm F. T.}\longrightarrow&
-{2 \langle {\bar d} d \rangle \over 3} A^u_G 
\gamma_5 \sigma^{\mu\nu} q_\mu p_\nu {1\over q^4} 
\nonumber \\
& \rightarrow& -{2 f_\pi \over 9} \langle {\bar d} d \rangle
\delta^2 
\gamma_5 \sigma^{\mu\nu} q_\mu p_\nu {1\over q^4}\ .
\label{ptope5}
\end{eqnarray}
Note in some cases, we have specified the steps before final 
expressions as they are useful for the extension to other couplings.
It should be remarked at this stage that, since we have taken into 
account the contributions from operators with two gluon lines, we 
also have to take into 
account the $\alpha_s$ correction in the leading term of the OPE.
However, that is a formidable task that will introduce only a small 
correction to our estimate and we will not consider it here.  
Rather, here we will concentrate on the effects from the higher 
twist component of the pion wave function.

The phenomenological side of this sum rule takes the form
\begin{eqnarray}
{g_{\pi N} \lambda_N^2  \over (q^2-m_N^2)^2} 
\gamma_5 \sigma^{\mu\nu} q_\mu p_\nu  + \cdot \cdot \cdot\ .
\end{eqnarray}
This expression is independent of the pseudoscalar-pseudovector coupling 
schemes and the ellipsis includes only the transitions
$N\rightarrow N^*$~\cite{hung1}.
Matching the two expressions and subsequent Borel transformation
lead to
\begin{eqnarray}
&&g_{\pi N}~ \lambda_N^2 (1+D_{\pi N} M^2)~e^{-m^2_N/M^2} \nonumber \\
&&=  
- {\langle {\bar q} q \rangle \over
f_\pi}  \left [ {M^4 E_0 (x) \over 12 \pi^2 }
+{4 \over 3 } f^2_\pi M^2  +
\left \langle {\alpha_s \over \pi} {\cal G}^2
\right \rangle
{1 \over 216 }
-{m_0^2 f^2_\pi \over 6 } + {26\over 27} f_\pi^2 \delta^2 
\right ]\label{ptsum}\ .
\end{eqnarray}
The contribution from $N\rightarrow N^*$ has been denoted
by $D_{\pi N}$.
The last term containing $\delta^2$ is new in this work, coming
from figures~\ref{fig1} (b) and (e).
This new term cancels the other term 
containing $m_0^2$ and makes the OPE larger.
Without this term, we previously reported $g_{\pi N} \sim 10$.
Thus, the new term should 
make better agreement with the empirical $\pi NN$ coupling.

\section{$\eta NN$ coupling and the $F/D$ ratio}
\label{sec:enn}

Extensions of the $\pi^0 pp$ sum rules to the $\eta NN$ coupling
is straightforward.  In this case, we consider the correlation
function with an $\eta$,
\begin{eqnarray}
i \int d^4 x e^{i q \cdot x} \langle 0 | T[J_p (x) 
{\bar J}_p (0)]| \eta (p) \rangle  \ .
\label{etwo}
\end{eqnarray}
In the SU(3) symmetric limit, there is no 
$\eta-\eta^\prime$ mixing and, within the OPE dimension
that we are considering, the strange quark
component of $\eta$ does not participate in the sum rule.
The difference from the $\pi^0 pp$ case is that, since $\eta$  
is isoscalar, the $d$-quark matrix element couples to
an $\eta$ with the same sign as the $u$-quark element does.
Thus, the quark-antiquark
elements with an $\eta$ contributing to our sum rules have
the following relations,
\begin{eqnarray}
\langle 0| {\bar u}(0) \gamma_\mu \gamma_5 u(x) | \eta (p)\rangle
&=&\langle 0| {\bar d}(0) \gamma_\mu \gamma_5 d(x) | \eta (p)\rangle
\nonumber \\
&\rightarrow& {i f_\eta \over \sqrt{3}}
\left [ p_\mu -{5\over 18} \delta^2  
\left ( {1\over 2} x^2 p_\mu  -{1\over 5} x_\mu x\cdot p \right )\right ]
\label{epv}\ ,\\
\langle 0 | {\bar u}(0) \gamma_5 \sigma_{\mu\nu}  u (x) | \eta (p) \rangle
&=&
\langle 0 | {\bar d}(0) \gamma_5 \sigma_{\mu\nu}  d (x) | \eta (p) \rangle
\nonumber \\
&\rightarrow& -i(p_\mu x_\nu - p_\nu x_\mu) 
{\langle{\bar u} u\rangle \over 6 \sqrt{3}f_\eta}\label{ept}\ ,\\
\langle 0| u(x)^\alpha_a i g_s ({\tilde G}^A (0))^{\sigma \rho} 
{\bar u}(0)^\beta_b | \eta(p) \rangle
&=&
\langle 0| d(x)^\alpha_a i g_s ({\tilde G}^A (0))^{\sigma \rho} 
{\bar d}(0)^\beta_b | \eta(p) \rangle\nonumber \\
&\rightarrow& -t^A_{ab} (\gamma_\theta)^{\alpha \beta} 
{f_\eta \delta^2 \over 48\sqrt{3}} 
(p^\rho g^{\theta \sigma} - p^\sigma g^{\theta \rho})
\label{ethree}\ .
\end{eqnarray}
Similarly for the $\pi^0 pp$ case, these matrix elements are the basic 
ingredients in calculating the OPE. In this construction,
the SU(3) breaking effects  are driven only
by $f_\eta \ne f_\pi$. 

Keeping in mind the sign change of the matrix elements involving
the $d$-quark, it is straightforward to 
obtain the OPE for the
$\eta pp$ coupling  directly from Eq.(\ref{pvope1}) - Eq.(\ref{pvope5})
for the $i\gamma_5 {\hat p}$ sum rule,
and from Eq.(\ref{ptope1}) - Eq.(\ref{ptope5}) for
the $\gamma_5 \sigma^{\mu\nu} q_\mu p_\nu$ sum rule.
For example in the case of figure~\ref{fig1} (a), we have
the similar expression as Eq.~(\ref{pvope1}) but now the $d$-quark
element adds up with the $u$-quark element to yield
the OPE
\begin{eqnarray}
{if_\eta \over 3 \sqrt{3} \pi^2} \gamma_5 {\hat p}
\left [q^2 ln(-q^2) + \delta^2 ln(-q^2) \right ]\ .
\end{eqnarray}
By similarly calculating for the other OPE, we obtain
the $i\gamma_5 {\hat p}$ sum rule,
\begin{eqnarray}
g_{\eta N}~m_N~\lambda_N^2 (1+&C_{\eta N}&M^2) e^{-m_N^2/M^2}
=
{f_\eta \over 3\sqrt{3} \pi^2} \left [ M^6E_1+
{5\over 2}\delta^2 M^4 E_0 \right ]
\nonumber \\
&+&{M^2 \over \sqrt{3}}\left [ {f_\eta \over 9}
\left \langle {\alpha_s \over \pi} {\cal G}^2 \right \rangle
+{2\over 9 f_\eta} \langle {\bar q} q\rangle^2  \right ]
- {2f_\eta \delta^2 \over 81\sqrt {3}}  
\left \langle {\alpha_s \over \pi} {\cal G}^2 \right \rangle 
+{m_0^2 \langle {\bar q} q\rangle^2 \over 36 \sqrt{3} f_\eta}\ .
\label{pvsumeta}
\end{eqnarray}
The RHS in the SU(3) limit ($f_\pi = f_\eta$)
should satisfy the SU(3) relation
for the $\eta NN$ coupling~\cite{swart},
\begin{eqnarray}
{g_{\eta N}\over g_{\pi N}}
\sim {1\over \sqrt{3}} (4\alpha -1)\ ,
\label{salpha}
\end{eqnarray}
where $\alpha = {F\over F+D}$.   
To identify the OPE corresponding to $\alpha$, we neglect the nucleon single
pole terms $C_{\pi N}$ and $C_{\eta N}$ for the time being.  A full analysis
including them will be given in later sections.  
By taking the ratio of  Eqs.(\ref{pvsum}) and (\ref{pvsumeta}) and
comparing with Eq.(\ref{salpha}),
we find a rather complicated expression for $\alpha$,
\begin{eqnarray}
4\alpha &\sim& \Bigg \{ {5 \over 6 \pi^2}M^6 E_1(x)
+{4\delta^2 \over 3\pi^2} M^4 E_0 (x) 
+ \left \langle {\alpha_s \over \pi} {\cal G}^2 \right \rangle
\left ( {7\over 36} M^2 -{11\over 324}\delta^2 \right )
+{4\over 9 f^2_\pi} \langle {\bar q} q\rangle^2 M^2  \nonumber \\
&&~~~ + {m_0^2 \langle {\bar q} q\rangle^2 \over 18 f^2_\pi}
\Bigg \}\nonumber \\
&\times& \Bigg \{
{1\over 2\pi^2} M^6 E_1 (x) + {\delta^2 \over 2\pi^2} M^4 E_0 (x) 
+  {1 \over 12}
\left \langle {\alpha_s \over \pi} {\cal G}^2 \right \rangle 
\left ( M^2 -{\delta^2\over 9}  \right )
+
{2\langle {\bar q} q \rangle^2  \over 9 f^2_\pi} M^2 \nonumber \\
&&~~~ +
{m_0^2\langle {\bar q} q \rangle^2 \over 36 f^2_\pi}
\Bigg \}^{-1}\ .
\label{apv}
\end{eqnarray}
Of course, this equation as is written here
should not be used to determine
the $F/D$ ratio because the important contribution
from the unknown nucleon single-pole terms has been neglected in
the construction. This equation only provides a 
consistency check for the OPE as the QCD side of the
sum rule should satisfy the SU(3) relations for the couplings.

By performing a similar calculation, we obtain for  
the $\gamma_5 \sigma^{\mu\nu} q_\mu p_\nu$ sum rule,
\begin{eqnarray}
&&g_{\eta N}~ \lambda_N^2 (1+D_{\eta N}M^2)~e^{-m^2_N/M^2} \nonumber \\
&&= 
{\langle {\bar q} q \rangle \over
\sqrt{3} f_\eta}  \left [ {M^4 E_0 (x) \over 12 \pi^2 }
-{4 \over 3 } f^2_\eta M^2  +
\left \langle {\alpha_s \over \pi} {\cal G}^2
\right \rangle
{1 \over 216 }
+{m_0^2 f^2_\eta \over 6 } - {26\over 27} f_\eta^2 \delta^2 
\right ]\label{ptsum2}\ .
\end{eqnarray}
Note the first and third terms, 
as they come from the OPE containing the $d$-quark elements
with $\eta$, have the opposite sign
from the corresponding terms in Eq.~(\ref{ptsum}).  
Because of this, there seems a delicate cancellation between
the OPE terms and the overall OPE strength becomes small.

Combining Eqs.~(\ref{ptsum}) and (\ref{ptsum2}) and
neglecting again the nucleon single-pole terms, $D_{\pi N}$
and $D_{\eta N}$, 
we identify $\alpha$ in the SU(3) limit ($f_\pi = f_\eta$),
\begin{eqnarray}
4\alpha &\sim& \Bigg \{ {8f_\pi \over 3 } M^2 
-{f_\pi m_0^2 \over 3}  
+ {52 f_\pi \delta^2 \over 27} \Bigg \} 
\nonumber \\
&\times& \Bigg \{ {M^4 E_0(x) \over 12 \pi^2 f_\pi}
+{4 f_\pi \over 3} M^2 + {26\over 27}f_\pi \delta^2
-{f_\pi m_0^2\over 6} + 
{1\over 216 f_\pi} 
\left \langle {\alpha_s \over \pi} {\cal G}^2 \right \rangle 
\Bigg \}^{-1} \ .
\label{apt}
\end{eqnarray}
The quark condensate is canceled in the ratio.
The expression is clearly different from Eq.~(\ref{apv}).
In our previous work, we obtain
from the $i\gamma_5$ sum rules beyond the chiral limit~\cite{hung6},
\begin{eqnarray}
4\alpha &\sim&  {4m_q m_0^2 \langle {\bar q} q \rangle^2 \over 3f_\pi} 
\nonumber \\
&\times& 
\Bigg \{ 
-m_\pi^2 M^4 E_0 (x) \left [
{\langle {\bar q}q \rangle \over 12 \pi^2 f_\pi}
              + {3 f_{3\pi} \over 4\sqrt{2}\pi^2} \right ]
+ {2 m_q\over f_\pi} \langle {\bar q}q \rangle^2 M^2
+ {m_\pi^2 \over 72 f_\pi} \langle {\bar q}q \rangle
\left \langle {\alpha_s \over \pi} {\cal G}^2\right \rangle\nonumber \\
&+&
{2m_q\over 3 f_\pi } m_0^2 \langle {\bar q}q \rangle^2
\Bigg \}^{-1}\ .
\label{aps}
\end{eqnarray}
Through the Gell-Mann$-$Oakes$-$Renner relation,
the $m_q$ as well as the $m_\pi^2$ dependence will be
canceled in the ratio. This expression 
is also not consistent with 
Eq.(\ref{apv}) or Eq.(\ref{apt}).
Therefore, depending on Dirac structures, we clearly have 
different expressions for $\alpha$. 
Later sections, we will discuss the numerical value of $\alpha$ when
we analyze the sum rules including the unknown single-pole terms.

\section{$\pi \Xi \Xi$, $\eta \Xi \Xi$, 
$\pi \Sigma \Sigma$ and $\eta \Sigma \Sigma$}
\label{sec:others}

Having identified the OPE corresponding to $\alpha$ in
the SU(3) limit, we apply, as consistency checks,
the sum rule framework introduced above
to the couplings, $\pi \Xi \Xi$, $\eta \Xi \Xi$,
$\pi \Sigma \Sigma$ and $\eta \Sigma \Sigma$.
In this extension, the strange quark enters to
the sum rules as an active degree of freedom.
One important constraint to be satisfied always is the SU(3) relations
for the couplings~\cite{swart}.  This means that the identification
of $\alpha$ made in Eq. (\ref{apv}) for the $i\gamma_5 {\hat p}$ sum rule
and in Eq.(\ref{apt}) for the $\gamma_5 \sigma^{\mu\nu} q_\mu p_\nu$
should be separately satisfied whenever SU(3) symmetry is
imposed on the OPE.  This statement is obvious because
the interpolating fields used for $\Xi$ and $\Sigma$ 
are constructed from the nucleon interpolating field via
the SU(3) rotation.  However, it is often the case that 
this SU(3) constraint is 
not properly imposed in QCD sum rule constructions 
of meson-baryon couplings in the SU(3) sector.  

One important limitation in this extension is 
related to quark mass corrections.  
Our sum rules are constructed at the order of 
${\cal O} (p_\mu)$ in the expansion of
the meson momentum.  The quark-mass terms are of higher 
orders and thus should not be included in this sum rule.
However, when the massive $s$-quark
is involved,
it is questionable whether the sum rules truncated at the
first order in $p_\mu$ is reliable:
the sum rule at the order $p_\mu$ may not 
properly represent
the total strength  of the correlators.
In this sense, our sum rules in this extension are 
somewhat limited and
a more systematic procedure which does not
rely on the expansion of the
meson momentum may be required for realistic prediction
for the couplings. Indeed,
the light-cone sum rule may be useful for that purpose but
in this case, QCD inputs contain the meson wave functions
at a specific point instead of their integrated
strength as in our case. Therefore, 
predictions may depend on a specific `ansatz' for the
wave functions~\cite{zhit}.  Moreover, there is an issue
in applying QCD duality in the construction of the light-cone
sum rule~\cite{hung4} and the usual application of 
QCD duality in QCD sum rules with external fields may
not be well satisfied~\cite{blok}.  
In future a more systematic method to overcome these difficulties
is anticipated.  
Nevertheless, our sum rules at the linear order in $p_\mu$
are  reliable as long as the SU(3) symmetric limit is imposed,
because, in that limit, the $s$-quark mass is small.   
Therefore, the discussion regarding the $F/D$ ratio is
reasonable.
When we discuss the couplings beyond the SU(3) limit, however,
this limitation should be noted.

We calculate the meson-baryon couplings for $\pi \Xi \Xi$, $\eta \Xi \Xi$,
$\pi \Sigma \Sigma$ and $\eta \Sigma \Sigma$ from a
correlation function of the type,
\begin{eqnarray}
i \int d^4 x e^{i q \cdot x} \langle 0 | T[J_{B} (x)
{\bar J}_{B} (0)]| M (p) \rangle \ ,
\label{mbtwo}
\end{eqnarray}
where $J_B$ is the corresponding baryon interpolating field 
and $|M(p) \rangle $ is the 
meson state of concern.  
For $\Xi$ and $\Sigma$, we use the interpolating fields~\cite{qsr}
\begin{eqnarray}
J_{\Xi} &=& 
-\epsilon_{abc} [ s_a^T C \gamma_\mu s_b ] \gamma_5 \gamma^\mu u_c\ ,
\nonumber\\ 
J_{\Sigma} &=&
\epsilon_{abc} [ u_a^T C \gamma_\mu u_b ] \gamma_5 \gamma^\mu s_c\ ,
\end{eqnarray}
respectively obtained from the nucleon interpolating field
via the SU(3) rotations. 
Calculation can be performed similarly as before. 
But, since the baryon interpolating fields have the
similar structure as the nucleon interpolating field, we can
easily obtain the OPE for each sum rule by making simple replacements
from the $\pi NN$ or $\eta NN$ sum rules. 
To be more specific, the OPE for the $\pi \Xi \Xi$ coupling is obtained
from that of the $\pi pp$ coupling by replacing the quark fields, 
$u \rightarrow s$ and $d \rightarrow u$.  The same replacements are
required to obtain the OPE for the $\eta \Xi \Xi$ from that of $\eta pp$.
For the $\pi \Sigma \Sigma$ and $\eta \Sigma \Sigma$ couplings, we need 
to replace $d \rightarrow s$ from the corresponding nucleon sum rules. 
A new ingredient
in this extension is the strange quark-antiquark component with the specific
meson of concern.  The strange quark-antiquark component does not couple to
a pion within the OPE dimension that we are considering. 
However, in the case of the $\eta$-baryon couplings,
there is nonzero strength between the strange quark-antiquark operators 
and an $\eta$. Its strength relative to the $u$ or $d$-quark
operators can be read off from the SU(3) Gell-Mann matrix.     
Namely, we have
\begin{eqnarray}
A_\mu^s (\eta) \equiv
\langle 0| {\bar s}(0) \gamma_\mu \gamma_5 s(x) | \eta (p)\rangle
&\rightarrow&- 
{2 \over \sqrt{3}} i f_\eta p_\mu + i{5\over 9\sqrt{3}} f_\eta \delta^2  
\left ( {1\over 2} x^2 p_\mu  -{1\over 5} x_\mu x\cdot p \right )
\label{spv2}\ ,\\
B_{\mu\nu}^s (\eta) \equiv
\langle 0 | {\bar s}(0) \gamma_5 \sigma_{\mu\nu}  s (x) | \eta (p) \rangle
&\rightarrow& + i(p_\mu x_\nu - p_\nu x_\mu) 
{\langle{\bar s} s\rangle \over 3 \sqrt{3} f_\eta}
\label{spt2}\ ,\\
\langle 0| s(x)^\alpha_a i g_s ({\tilde G}^A (0))^{\sigma \rho} 
{\bar s}(0)^\beta_b | \eta(p) \rangle
&\rightarrow & 
+ t^A_{ab} (\gamma_\theta)^{\alpha \beta} {f_\eta \delta^2 \over 24\sqrt{3}} 
(p^\rho g^{\theta \sigma} - p^\sigma g^{\theta \rho})\ .
\label{sthree}
\end{eqnarray}
Compared with Eqs.(\ref{pv2}), (\ref{pt}),(\ref{three}),
the strange quark-antiquark elements with an $\eta$ have
the overall sign consistent with the d-quark components with
a pion but the magnitude has been multiplied by the factor $2/\sqrt{3}$
as it should be.
In Eqs.~(\ref{spv2}) and (\ref{sthree}), 
the SU(3) breaking is reflected 
only in $f_\eta$ 
but in Eq.~(\ref{spt2}), there is another SU(3) breaking source,
the strange quark condensate.   
As we know how to go to the SU(3) symmetric limit from these 
two breaking sources, the SU(3) relations for the couplings can be
easily investigated in this approach.

With these differences in mind, we can straightforwardly calculate the
OPE for each coupling. In the case of the $i\gamma_5 {\hat p}$ Dirac
structure,
we obtain the $\pi \Xi \Xi$ sum rule,
\begin{eqnarray}
g_{\pi \Xi}~m_{\Xi}~ \lambda_{\Xi}^2 (1+C_{\pi \Xi} M^2)~e^{-m^2_{\Xi}/M^2}
&=&  
-{f_\pi\over 12\pi^2} \left [ M^6 E_1  - 2\delta^2 M^4 E_0  \right ]
\nonumber \\
&+& {f_\pi \over 72}
\left \langle {\alpha_s \over \pi} {\cal G}^2 \right \rangle 
\left ( M^2  - {5 \delta^2 \over 9} \right )\ .
\label{pxipvsum} 
\end{eqnarray}
Again, neglecting the unknown single pole term $C_{\pi \Xi}$, it is easy 
to see that in the SU(3) limit the RHS satisfies the SU(3) relation,
\begin{eqnarray}
{g_{\pi \Xi} \over g_{\pi N}} \sim 2 \alpha -1\ ,
\label{su3rel1}
\end{eqnarray}
if we identify $\alpha$ as Eq.~(\ref{apv}).  This suggests
that our approach makes sense at least in retrieving 
consistently the SU(3) relation for the coupling.
For the other couplings, we similarly proceed the calculations. 
The LHS side
of each sum rule has the similar structure as that of Eq.(\ref{pxipvsum})
but now we have different baryon mass, the coupling,
and the strength of the interpolating field to the baryon of
concern.
The RHS of each sum rule for the $i\gamma_5 {\hat p}$ structure
is obtained as follows 
\begin{eqnarray}
&\eta \Xi \Xi& :~~
-{1\over \sqrt{3}} \Bigg \{ {11 f_\eta \over 12 \pi^2}M^6 E_1
+ {7f_\eta  \delta^2 \over 6 \pi^2} M^4 E_0 
+ {f_\eta \over 72}
\left \langle {\alpha_s \over \pi} {\cal G}^2 \right \rangle
\left ( 13 M^2  - {17 \delta^2 \over 9} \right )
\nonumber \\
&~~~~~~~~~~+& {4\langle {\bar s} s \rangle^2 \over 9 f_\eta} M^2 
+ {m_0^2 \langle {\bar s} s \rangle^2 \over 18 f_\eta} \Bigg \}
\label{etaxipvsum} \ ,\\
&\pi \Sigma \Sigma& :~~ 
{f_\pi\over 12\pi^2} \left [ 5M^6 E_1  + 8\delta^2 M^4 E_0  \right ] 
+ {f_\pi \over 72}
\left \langle {\alpha_s \over \pi} {\cal G}^2 \right \rangle
\left ( 7 M^2  - {11 \delta^2 \over 9} \right )
\nonumber \\
&~~~~~~~~~~+& {2\langle {\bar q} q \rangle^2 \over 9 f_\pi} M^2 
+ {m_0^2 \langle {\bar q} q \rangle^2 \over 36 f_\pi}  
\label{psigpvsum}\ ,\\
&\eta \Sigma \Sigma& :~~
{1\over \sqrt{3}} \Bigg \{ {7 f_\eta \over 12 \pi^2}M^6 E_1
+ {f_\eta  \delta^2 \over 3 \pi^2} M^4 E_0 
+ {f_\eta \over 72}
\left \langle {\alpha_s \over \pi} {\cal G}^2 \right \rangle
\left ( 5 M^2  - {\delta^2 \over 9} \right )
\nonumber \\
&~~~~~~~~~~+& {2\langle {\bar q} q \rangle^2 \over 9 f_\eta} M^2 
+ {m_0^2 \langle {\bar q} q \rangle^2 \over 36 f_\eta} \Bigg \}\ .
\label{etasigpvsum}
\end{eqnarray}
Again it is straightforward to show that in the SU(3) limit 
($f_\pi = f_\eta$, $\langle {\bar s} s \rangle =\langle {\bar q} q \rangle$
and $\lambda_N=\lambda_\Xi=\lambda_\Sigma$) the RHS 
of each sum rule
satisfies the SU(3) relations for the couplings,
\begin{eqnarray}
{g_{\eta \Xi}\over g_{\pi N}} \sim -{1\over \sqrt{3}} (1 + 2 \alpha)
\;; \quad
{g_{\pi \Sigma}\over g_{\pi N}} \sim 2 \alpha
\;; \quad
{g_{\eta \Sigma}\over g_{\pi N}} \sim {2\over \sqrt{3}}(1-\alpha)\ ,
\label{su3rel2}
\end{eqnarray}
with $\alpha$ given in Eq.~(\ref{apv}). 
Therefore the consistency check has been made for these sum rules.

In the case of the $\gamma_5 \sigma^{\mu\nu} q_\mu p_\nu$ Dirac
structure, the LHS side of the sum rule for meson($m_i= \pi~{\rm or}~
\eta$)-baryon ($B_j = \Xi~{\rm or}~\Sigma$) coupling 
takes the form
\begin{eqnarray}
g_{m_i B_j}~ \lambda_{B_j}^2 
(1+D_{m_i B_j} M^2)~e^{-m^2_{B_j}/M^2}\ .
\end{eqnarray}
In the RHS, we have the following set of the sum rules,
\begin{eqnarray}
&\pi \Xi \Xi:~~& 
{\langle {\bar q} q \rangle \over 12 \pi^2 f_\pi} M^4 E_0 + 
{\langle {\bar q} q \rangle \over 216 f_\pi}
\left \langle {\alpha_s \over \pi} {\cal G}^2 \right \rangle
\label{pixiptsum}\ ,\\
&\eta \Xi \Xi:~~&
{\langle {\bar q} q \rangle \over \sqrt{3}}
\Bigg \{ {M^4 E_0 \over 12 \pi^2 f_\eta} + {8 f_\eta M^2 \over 3}
+{52 f_\eta \delta^2 \over 27} -{f_\eta m_0^2 \over 3}
+ {1\over 216 f_\eta} 
\left \langle {\alpha_s \over \pi} {\cal G}^2 \right \rangle
\Bigg \}
\label{etaxiptsum}\ ,\\
&\pi \Sigma \Sigma:~~&
\langle {\bar s} s \rangle \Bigg \{ 
-{4f_\pi \over 3}M^2 - {26f_\pi \delta^2 \over 27} + {f_\pi m_0^2 \over 6}
\Bigg \} 
\label{pisigptsum}\ ,\\
&\eta \Sigma \Sigma:~~&
{\langle {\bar s} s \rangle \over \sqrt{3}} \Bigg \{
-{M^4 E_0 \over 6 \pi^2 f_\eta} - {4 f_\eta \over 3} M^2
-{26 f_\eta \delta^2 \over 27} - 
{1\over 108 f_\eta}
\left \langle {\alpha_s \over \pi} {\cal G}^2 \right \rangle
+{f_\eta m_0^2 \over 6} 
\Bigg \}
\label{etasigptsum}\ .
\end{eqnarray}
It can be easily checked that
these RHS of the sum rules satisfy the SU(3) relations for the
couplings Eqs.~(\ref{su3rel1}) and (\ref{su3rel2}) if we identify
$\alpha$ as given in Eq.(\ref{apt}), again making sure the
consistency with SU(3) symmetry.

\section{Analysis in the SU(3) symmetric limit} 
\label{sec:su3}

We now analyze the sum rules provided in the previous sections
within the SU(3) symmetric limit. What we have demonstrated so far
is that we have different identifications for the $F/D$ ratio
depending on the Dirac structure considered.
Each set of sum rules satisfies the SU(3) relations for the
couplings with different identifications for the $F/D$ ratio.
In this section, we calculate numerical values of 
the $F/D$ ratio from the $i\gamma_5 {\hat p}$ and 
$\gamma_5 \sigma_{\mu\nu} q^\mu p^\nu$
sum rules separately, and see if they are consistent with the
previous calculation beyond the chiral limit~\cite{hung6}. 
In the SU(3) limit, all the baryon masses are the same
$m_N=m_\Xi=m_\Sigma$, and we also have the relations,
$\langle {\bar s}s \rangle
=\langle {\bar q}q \rangle$ and $f_\eta = f_\pi$.
Moreover, from the baryon mass sum rules~\cite{qsr}, the strength
of each interpolating field to the low-lying baryon should be
the same in this limit, $\lambda_N=\lambda_\Xi=\lambda_\Sigma$.
We take the conventional values for the QCD parameters in this analysis,
\begin{eqnarray}
\langle {\bar q} q \rangle &=& -(0.23~ {\rm GeV})^3\;; \quad
\left \langle {\alpha_s \over \pi} {\cal G}^2\right \rangle
= (0.33~{\rm GeV})^4
\nonumber \\
\delta^2&=&0.2~{\rm GeV}^2\;; \quad m_0^2=0.8~{\rm GeV}^2\ .
\end{eqnarray}

We start with the $i\gamma_5 {\hat p}$ sum rules in the SU(3) limit.
This structure for the $\pi NN$ coupling has been studied 
in Ref.~\cite{hung2}. 
By revisiting them here, we want to see 
whether the higher dimensional
operators included in this work change the previous results.
To proceed, we rearrange the sum rule equations for the
$i\gamma_5 {\hat p}$ structure, Eqs. (\ref{pvsum}),(\ref{pvsumeta}),
(\ref{pxipvsum}),(\ref{etaxipvsum}), (\ref{psigpvsum}),(\ref{etasigpvsum}) 
into the form,
\begin{eqnarray}
a + b M^2 = f(M^2)\ ,
\label{form}
\end{eqnarray}
by transferring baryon masses and the exponential factors to
the RHS of the sum rules. Thus, in the case of the
$\pi NN$ sum rule, Eq. (\ref{pvsum}), the parameters represent that
\begin{eqnarray}
a=g_{\pi N} \lambda_N^2\;; \quad
b=g_{\pi N} \lambda_N^2 C_{\pi N}\ ,
\end{eqnarray} 
and similarly for the other couplings.

\begin{figure}
\caption{ The Borel mass dependence of $a+bM^2$ for the $i\gamma_5 {\hat p}$
sum rules in the SU(3) symmetric limit.
The continuum threshold $S_0 = 2.07 $ GeV$^2$, corresponding
to the Roper resonance, is used to obtain the solid lines.
To see the sensitivity to the continuum threshold,
the dashed-lines with $S_0 =2.57 $ GeV$^2$ are also
plotted.
}
\label{fig2}

\setlength{\textwidth}{6.1in}   
\setlength{\textheight}{9.in}  
\centerline{%
\vbox to 2.4in{\vss
   \hbox to 3.3in{\includegraphics{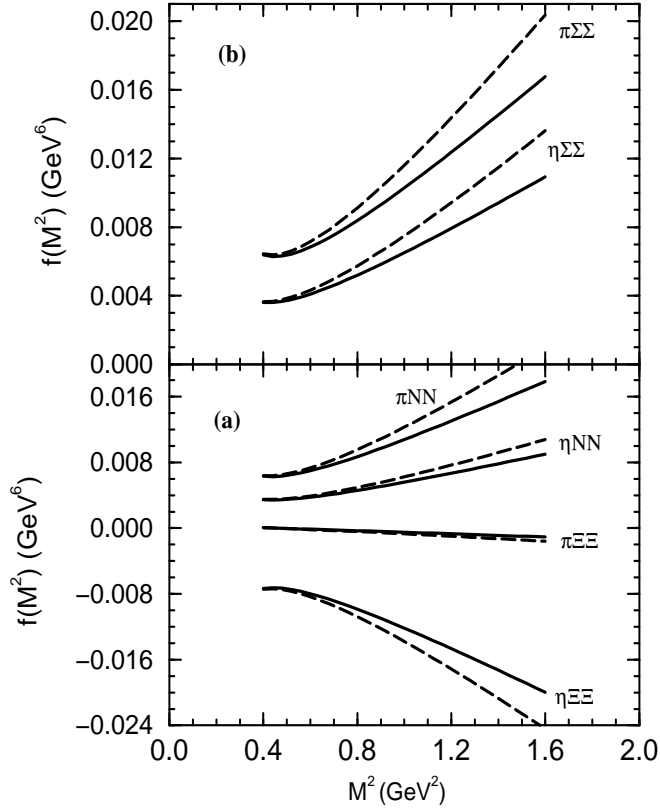}\hss}}
}
\vspace{130pt}
\end{figure}

In figures~\ref{fig2} (a) (b), we plot the
RHS of the sum rules $f(M^2)$ for the couplings. 
The continuum threshold is set to $S_0 =2.07$ GeV$^2$ corresponding
to the Roper resonance  and is used in obtaining
the solid lines.  To see the sensitivity to the continuum
threshold, the Borel curves  with $S_0 =2.57$ GeV$^2$ are also
shown with the dashed lines.
The $\pi NN$ Borel curves are almost the same as the ones 
presented in Ref.~\cite{hung2}
indicating that the higher dimensional operators included in this work
do not change the previous results.  
By linearly fitting each Borel curve within an appropriate Borel
window, we extract the two phenomenological parameters
$a$ and $b$. The parameter $a$ is given by the intersection of the
vertical axis ($M^2=0$) with the best fitting straight line,
and the parameter $b$ is given by the slope of the line. 
As one can see from the figures, in most sum rules, there is huge
sensitivity to the
continuum threshold, which prevents us to extract reliably the parameters 
of the concern.  At $M^2\sim 1$ GeV$^2$,
the $\pi NN$ Borel curve undergoes 14 \% change due to the 
continuum parameter, which however yields rather different value of
$a$ as shown in table~\ref{tab1}.
For the other sum rules, we can also see from the table that
the extracted parameters are highly  sensitive to
the continuum. 
One of the reasons may be, as suggested in Ref.~\cite{hung2},
because higher resonances with different parities
add up in forming the continuum or the unknown single pole terms.
The SU(3) parameter $\alpha = F/(F+D)$
extracted from table~\ref{tab1} is $\alpha \sim 1.24$ when the
continuum parameter $S_0 = 2.07$ GeV$^2$ is used.  This
gives $F/D=-5.17$. 
But with $S_0 = 2.57$ GeV$^2$, we have totally different value,
$\alpha = 0.288$, which yields $F/D=0.4$.  Therefore, the 
$i\gamma_5 {\hat p}$ sum rules
may not be useful in determining the $F/D$ ratio.

Figure~\ref{fig3} shows the Borel curves for the Dirac structure 
$\gamma_5 \sigma_{\mu\nu} q^\mu p^\nu$. 
The sum rules, Eqs. (\ref{ptsum}), 
(\ref{ptsum2}), (\ref{pixiptsum}), (\ref{etaxiptsum}),(\ref{pisigptsum})
and (\ref{etasigptsum})
are arranged into the form of Eq.~(\ref{form}) and the
RHS of that  is plotted for each coupling in the figure.
Hence, the best fitting straight line
within a Borel window will provide us with the 
parameters $a$ and $b$, which represent the same quantities as before.
In contrast to the $i\gamma_5 {\hat p}$ sum rules, the Borel curves in
this case are rather
insensitive to the continuum parameter $S_0$. The sensitivity of the Borel 
curves to the continuum
at  $M^2\sim 1 $ GeV$^2$ is about 2 \% level.
Therefore, this $\gamma_5 \sigma_{\mu\nu}q^\mu p^\nu$
structure may provide a useful constraint for the $F/D$ ratio.

The $\pi NN$ sum rule in this work has been improved from the previous
calculations of Ref.~\cite{hung1,hung2} by including gluonic contributions 
combined with the external pion state. They are from the
three-particle pion wave functions, which produces
the term involving $\delta^2$ in Eq.~(\ref{ptsum}).
The new term appears in the highest dimension and 
cancels the other OPE at the same dimension containing 
the quark-gluon mixed parameter $m_0^2$. 
Thus, the total OPE is well saturated by the first two OPE terms.
(Note that the term involving the gluon condensate in the
same dimension contributes negligibly to the total OPE.)
Combining Eq.~(\ref{ptsum}) with the chiral-odd nucleon sum rule 
and taking the standard sum rule analysis, we obtain,
\begin{eqnarray}
g_{\pi N} \sim 13-14\ .
\label{ptour}
\end{eqnarray}
The errors are coming from how we choose the Borel window.
This is certainly consistent with its empirical value as well as
the one obtained from the sum rule beyond the chiral limit~\cite{hung5}.
This also means that the gluonic contributions which were not
included in our previous study~\cite{hung1,hung2} are important
in stabilizing the sum rules and in obtaining
the coupling agreeing with the phenomenology.  

Other Borel curves for the sum rules,
Eqs.~(\ref{ptsum2}), (\ref{pixiptsum}), (\ref{etaxiptsum}),
(\ref{pisigptsum})
and (\ref{etasigptsum}) are plotted in figure~\ref{fig3}.  As we 
have demonstrated, 
each OPE satisfies the SU(3) relation if we identify 
$\alpha$ as given in Eq.~(\ref{apt}).   Thus, as far
as the OPE is concerned, all the sum rules in the SU(3) limit are related
by the SU(3) rotations.  This means that the same Borel window should
to be used for the other couplings.  Table~\ref{tab2} shows
our results from the $\gamma_5 \sigma_{\mu\nu} q^\mu p^\nu$ sum rules.
In the $\pi \Sigma \Sigma$ case, there is no dependence on
the continuum threshold.
The ratios given in the fourth column are directly related to the
SU(3) relations for the couplings.  From them, we 
consistently obtain $\alpha=0.44$, which yields 
\begin{eqnarray}
F/D \sim 0.78\ .
\end{eqnarray}
This is a factor of 4 larger than our previous value $F/D\sim 0.2$ 
determined beyond the chiral limit~\cite{hung6}, clearly
indicating the Dirac structure dependence of a sum rule result. 
It is even
larger than the value from the SU(6) symmetry $F/D \sim 2/3$ or
the recent value $F/D \sim 0.57$~\cite{ratcliffe}.  
Using the empirical value for the $\pi NN$ coupling, $g_{\pi N}=13.4$,
we obtain the following couplings in the SU(3) limit (indicated
by the superscript below),
\begin{eqnarray}
g^{(S)}_{\eta N} &=& 5.76\;; \quad
g^{(S)}_{\pi \Xi} = -1.61 \;; \quad 
g^{(S)}_{\eta \Xi} = -14.47\ ,\nonumber \\
g^{(S)}_{\pi \Sigma} &=& 11.79 \;; \quad
g^{(S)}_{\eta \Sigma} = 8.58\ .
\end{eqnarray}
These values are in contrast with the ones determined
beyond the chiral limit\cite{hung6},
\begin{eqnarray}
g^{(S)}_{\eta N} &=& -2.3\;; \quad
g^{(S)}_{\pi \Xi} = -8.7 \;; \quad
g^{(S)}_{\eta \Xi} = -10.5\ ,\nonumber \\
g^{(S)}_{\pi \Sigma} &=& 4.7 \;; \quad
g^{(S)}_{\eta \Sigma} = 12.8\ .
\end{eqnarray}
Once again, the Dirac structure
dependence~\cite{hung2} of a sum rule result is clearly exhibited. 

What could be the reasons for this Dirac structure dependence ?
Since the $i\gamma_5$ structure has the same chirality as
the $\gamma_5 \sigma_{\mu\nu}q^\mu p^\nu$ structure, the dependence
can not be attributed only to the higher resonance
contributions~\cite{hung2}. 
It seems rather due to the
use of Ioffe current for baryon interpolating fields. 
In fact, Ioffe current is a specific choice for the nucleon
interpolating field from a more general current of the type, 
\begin{eqnarray}
J_N=2 \epsilon_{abc}[(u_a^{\rm T} C d_b) \gamma_5 u_c
+t (u_a^{\rm T} C \gamma_5 d_b) u_c]\ .
\label{nucl1}
\end{eqnarray}
That is, when $t=-1$, this current reduces to Ioffe current.
One speculation that we can think of is that the choice with $t=-1$ is 
not optimal for the nucleon. 
Other speculation is the following. Since we obtain
the right strength for the $\pi NN$ coupling  at 
least from the $i\gamma_5$ and $\gamma_5 \sigma_{\mu\nu}q^\mu p^\nu$ sum 
rules, it could be
that  the Ioffe current is fine for the
nucleon but its SU(3) rotated versions may not be optimal
for the hyperons.   Further studies are necessary to
understand this problem in future.

\section{Meson-baryon couplings from the pseudotensor structure} 
\label{sec:bsu3}

In the previous section, we studied the SU(3) relations
for the couplings in the SU(3) limit from the 
$i\gamma_5 {\hat p}$ and $\gamma_5 \sigma_{\mu\nu} q^\mu p^\nu$
sum rules respectively.
The $i\gamma_5 {\hat p}$ sum rules contain strong
dependence on the continuum parameter and may not be
relevant for our purpose of obtaining the $F/D$ ratio. 
On the other hand,
the $\gamma_5 \sigma_{\mu\nu} q^\mu p^\nu$ sum rules 
have been found to provide a reasonable $\pi NN$ coupling
with less sensitivity to the continuum threshold.
The obtained valued for the $F/D$ ratio does not however 
agree with the previous result beyond the chiral limit.
Thus, the $\gamma_5 \sigma_{\mu\nu} q^\mu p^\nu$ sum rules
provide another set of couplings when the calculation
is performed beyond the SU(3) limit.
As we have emphasized, however, our results for 
strangeness baryons should be taken with some caution 
because our sum rules constructed at the order 
${\cal O} (p_\mu)$ may draw a doubt 
when the strange quark is involved.  
The strange quark mass, as it is higher than ${\cal O} (p_\mu)$,
should not be included in our approach.  A question therefore
is whether or not the sum rules truncated at order of ${\cal O} (p_\mu)$
make sense when the massive strange quark is involved.
A more systematic method may be needed in future to verify (or
refute) our approach here.  But in our standpoint, there
is no such a method at present.

With this limitation in mind, we present our results
for the $\gamma_5 \sigma_{\mu\nu}q^\mu p^\nu$ sum rules
beyond the SU(3) limit.
In our analysis, we will use the SU(3) breaking parameters
\begin{eqnarray}
f_\eta =1.2 f_\pi\;; \quad
\langle {\bar s} s \rangle = 0.8 \langle {\bar q} q \rangle\ .
\end{eqnarray}
The $f_\eta$ value is from Ref.~\cite{feldmann}.
In addition, phenomenological parameters such as
baryon masses and the strengths $\lambda_{B_j}$ 
will change as we move away from the SU(3) limit.
We take the empirical values for the masses and, for 
the strengths, we will discuss them as we move along. 
We ignore the SU(3) breaking driven by 
the singlet-octet mixing.

Figure~\ref{fig4} shows the Borel curves for the
$\pi \Sigma \Sigma$, $\eta \Sigma \Sigma$,
$\pi \Xi \Xi$ and $\eta \Xi \Xi$.  
Around the resonance masses ($M^2\sim 1.41$ GeV$^2$ for $\Sigma$
and $M^2\sim 1.73$ GeV$^2$ for $\Xi$), they are
well-fitted by a straight line,   
suggesting that the dependence on the chosen Borel window
is marginal.   As we have discussed, the dependence on the
continuum threshold is also small.  The $\eta NN$ curve
(not shown) basically has the similar features as the
case in the SU(3) symmetric limit but is shifted up slightly.
The best fitting parameters are given in table~\ref{tab3}.
Due to the unknown strengths $\lambda_{B_j}$ ($B_j=N,\Sigma,\Xi$),
we here present ratios of the couplings obtained from table~\ref{tab3},
\begin{eqnarray}
{g^{(B)}_{\eta N}\over g^{(B)}_{\pi N}} &=& 0.55\;; \quad
{g^{(B)}_{\eta \Xi} \over g^{(B)}_{\pi \Xi}} = 4.2\;; \quad 
{g^{(B)}_{\eta \Sigma} \over g^{(B)}_{\pi \Sigma}} = 0.94\ .
\end{eqnarray}
In comparison with the ratios in the SU(3) limit,
\begin{eqnarray}
{g^{(S)}_{\eta N}\over g^{(S)}_{\pi N}} &=& 0.43\;; \quad
{g^{(S)}_{\eta \Xi} \over g^{(S)}_{\pi \Xi}} = 8.9\;; \quad 
{g^{(S)}_{\eta \Sigma} \over g^{(S)}_{\pi \Sigma}} = 0.74\ .
\end{eqnarray}
the SU(3) breaking effects are huge for the meson-$\Xi$ couplings
but not so large for the other couplings.

Let's us now compare our results to that of other works.
In table~\ref{tab4}, our results in the
SU(3) limit and that from Refs.~\cite{rijken1,bh}
are shown.
The couplings in Ref.~\cite{rijken1} are based on the assumption of
the hyperon-nucleon potentials obeying SU(3) symmetry.   
Except for the results on $g_{\pi \Xi}$ and
$g_{\eta \Sigma}$, our results qualitatively agree with that of 
Ref.~\cite{rijken1}. Another approach based on the QCD 
parametrization method~\cite{bh} gives results (see the 5th column of
table~\ref{tab4}.) not so different from Ref.~\cite{rijken1}. 
Comparing our results in SU(3) to that in Ref.~\cite{bh},
we find qualitative agreement for $\pi \Sigma \Sigma$ but
large discrepancy in $\pi \Xi \Xi$.
To see the SU(3) breaking directly reflected in the couplings,
we simply use the scaling proposed by Ref.~\cite{dey},
$\lambda^2_{B_j}\propto m^6_{B_j}$, and calculate each  
coupling.  As shown in the 3rd column of table~\ref{tab4}, most couplings
change noticeably as we turn on the SU(3) breaking effects.
As a result, they do not agree with the ones from other works.
However, our results with the broken SU(3) should be taken with
cautions.  Since our sum rules are constructed at the order $p_\mu$,
the couplings obtained are the ones  at the kinematical point $p^2=0$.
But the physical couplings are defined 
at the kinematical point  $p^2=m^2_{\pi , \eta}$.  Therefore,
in the $\eta BB$ cases, one can expect some changes in this extrapolation.
Furthermore our formalism should be improved by including higher 
effects of the broken SU(3)
more systematically.

\section{Summary}
\label{sec:sum}

In this work, we have developed QCD sum rules beyond
the soft-meson limit for the diagonal meson-baryon couplings,
$\pi NN$, $\eta NN$, $\pi \Xi \Xi$, $\eta \Xi \Xi$,
$\pi \Sigma \Sigma$ and $\eta \Sigma \Sigma$.
The Dirac structures $i\gamma_5 {\hat p}$ and
$\gamma_5 \sigma_{\mu\nu} q^\mu p^\nu$  are separately considered in
constructing the sum rules.
In the first stage, we have improved the previous calculations of
the $\pi NN$ coupling by including three-particle pion wave functions
mediated by the gluonic tensor.  The $\gamma_5 \sigma_{\mu\nu} q^\mu p^\nu$
sum rule in this revision provides the $\pi NN$ coupling closed to 
its empirical value, while no critical change has been observed for the
$i\gamma_5 {\hat p}$ sum rules.  
By extend the $\pi NN$ sum rules to the other couplings,
we have studied the SU(3) relations for the couplings.
Depending on the Dirac structure considered,
we have reported different identifications of the $F/D$
ratio.
Therefore, our findings support the previous claim of the
Dirac structure dependence of a sum rule result~\cite{hung2}.
In the sum rule analysis,
the  $i\gamma_5 {\hat p}$  sum rules were found to give
the results very sensitive to the continuum threshold.
On the other hand, stable results are obtained from the
$\gamma_5 \sigma_{\mu\nu} q^\mu p^\nu$  sum rule, which however
is not consistent with the previous results obtained from
the sum rule beyond the chiral limit~\cite{hung6}. 
We have therefore provided a different set of the couplings in
the SU(3) limit and beyond the SU(3) limit using the
Dirac structure $\gamma_5 \sigma_{\mu\nu} q^\mu p^\nu$.
The obtained $F/D$ ratio from the 
$\gamma_5 \sigma_{\mu\nu} q^\mu p^\nu$  sum rules is $0.78$, slightly 
larger than the SU(6) value of $2/3$.  
We have also discussed the SU(3) breaking effects in the
couplings.

\acknowledgments
This work is supported in part by the 
Grant-in-Aid for JSPS fellow, and
the Grant-in-Aid for scientific
research (C) (2) 11640261 
of  the Ministry of Education, Science, Sports and Culture of Japan.
The work of  H. Kim is also supported by Research Fellowships of
the Japan Society for the Promotion of Science.
The work of S. H. Lee is supported by  the 
KOSEF grant number 1999-2-111-005-5 and by the BK 21 project of the
Korean Ministry of Education.

\begin{appendix}

\section{Derivation of the pseudovector matrix elements}
\label{sec:appa}

Here we evaluate the matrix elements
given in Eqs.~(\ref{pv1}) and (\ref{pv2}) by considering
\begin{eqnarray}
\langle 0| {\bar d}(0) \gamma_\mu \gamma_5 u(x) | \pi^+ (p)\rangle
\label{axial}
\end{eqnarray}
to leading  order in the pion momentum $p_\mu$.
We consider the case with a charged pion to use some informations
from Ref.~\cite{krippa} for twist-4 element. 
The matrix elements for the neutral pion, which are of our concern, can be 
obtained simply by isospin rotations afterward.
By expanding $u(x)$ in $x_\mu$, we have
\begin{eqnarray}
u(x) = u(0) + x_\mu \partial^\mu u(0) +
{1\over 2} x_\mu x_\nu \partial^\mu \partial^\nu u(0) + \cdot \cdot \cdot\ .
\end{eqnarray}
In the fixed-point gauge ($x_\mu A^\mu =0$), the partial derivative can be
replaced by the covariant derivative, $\partial_\mu \rightarrow
D_\mu \equiv \partial_\mu -ig_s A_\mu $.
In this expansion of Eq.(\ref{axial}), 
the first term is given by the
PCAC,
\begin{eqnarray}
\langle 0| {\bar d}(0) \gamma_\mu \gamma_5 u(0) | \pi^+ (p)\rangle
=i\sqrt{2} f_\pi p_\mu\ ,
\end{eqnarray}
where $f_\pi =93$ MeV.
The second term in the expansion, as it contains one covariant derivative,
is linear in the quark mass, ${\cal O}(m_q)$, which is higher order 
than $p_\mu$.   
The third term in the expansion 
containing two covariant derivatives can be written
\begin{eqnarray}
&&{1\over 2} x^\alpha x^\beta \langle 0| {\bar d}(0) \gamma_\mu \gamma_5
D_\alpha D_\beta  u(0) | \pi^+(p)\rangle\nonumber \\
&&~~~~=
{1\over 4} x^\alpha x^\beta \langle 0| {\bar d}(0) \gamma_\mu \gamma_5
(D_\alpha D_\beta + D_\beta D_\alpha) u(0) | \pi^+(p)\rangle\ .
\label{second}
\end{eqnarray}
The matrix element in the RHS sandwiched by the vacuum and the pion state
is symmetric under $ \alpha \leftrightarrow \beta$. Hence, it
should be built by symmetrically combining the available four vector 
$p_\mu$ and the metric $g_{\mu \nu}$. 
At the first order in $p_\mu$, it is easy to see that 
\begin{eqnarray}
\langle 0| {\bar d}(0) \gamma_\mu \gamma_5
{1\over 2} (D_\alpha D_\beta + D_\beta D_\alpha) u(0) | \pi^+(p)\rangle
=g_{\alpha \beta} p_\mu B + 
(g_{\alpha \mu} p_\beta + g_{\beta \mu} p_\alpha) C \ .
\label{twist4}
\end{eqnarray}
Note, other possible combinations are higher order than $p_\mu$.
To determine the invariant functions $B$ and $C$,
first let us multiply $g_{\alpha \beta}$ on both sides of Eq.~(\ref{twist4})
to obtain,
\begin{eqnarray}
\langle 0| {\bar d}(0) \gamma_\mu \gamma_5
D^2 u(0) | \pi^+(p)\rangle
= (4 B + 2 C) p_\mu\ .
\label{eq1}
\end{eqnarray} 
Other constraint equation can be  obtained by multiplying $g_{\alpha \mu}$
on Eq.~(\ref{twist4}),
\begin{eqnarray}
\langle 0| {\bar d}(0) \gamma_5
{1\over 2} (- {\hat D}  D_\beta - D_\beta {\hat D}) u(0) | \pi^+(p)\rangle
= (B + 5C) p_\beta \ ,
\label{a7}
\end{eqnarray}
where ${\hat D} \equiv \gamma_\mu D^\mu$.
The LHS is linear in the quark mass $m_q$ and higher order in
chiral counting than
the order $p_\mu$.  Thus, to leading order in $p_\mu$, the
LHS of Eq~(\ref{a7}) should be zero, which yields the relation,
\begin{eqnarray}
B=-5C\ .
\label{eq2}
\end{eqnarray}
Combining this with Eq.~(\ref{eq1}), we have 
\begin{eqnarray}
B p_\mu &=& {5\over 18} \langle 0| {\bar d}(0) \gamma_\mu \gamma_5
D^2 u(0) | \pi^+(p)\rangle \nonumber\ ,\\
C p_\mu &=& -{1\over 18} \langle 0| {\bar d}(0) \gamma_\mu \gamma_5
D^2 u(0) | \pi^+(p)\rangle \ .
\end{eqnarray}
According to Ref.~\cite{krippa}, the matrix element in
the RHS is given by 
\begin{equation}
\langle 0| {\bar d}(0) \gamma_\mu \gamma_5 D^2 u(0) | \pi^+(p)\rangle =
-i\sqrt{2} f_\pi \delta^2 p_\mu\ ,
\label{bkt4}
\end{equation} 
with $\delta^2=0.2$ GeV$^2$.
Therefore, we have
\begin{eqnarray}
B= -i{5\sqrt{2}\over 18} f_\pi \delta^2\;; \quad
C= i{\sqrt{2}\over 18} f_\pi \delta^2\ .
\end{eqnarray}
Using this result in Eq.~(\ref{twist4}) and putting into
Eq.~(\ref{second}), we obtain 
\begin{eqnarray}
{1\over 2} x^\alpha x^\beta \langle 0| {\bar d}(0) \gamma_\mu \gamma_5
D_\alpha D_\beta  u(0) | \pi^+(p)\rangle
= -i{5\over 18} \sqrt{2} f_\pi \delta^2 
\left ( {1\over 2} x^2 p_\mu  -{1\over 5} x_\mu x\cdot p \right )\ .
\end{eqnarray}
Thus, up to twist-4 but to leading order 
in $p_\mu$, we have the expansion
\begin{eqnarray}
\langle 0| {\bar d}(0) \gamma_\mu \gamma_5 u(x) | \pi^+ (p)\rangle
= i\sqrt{2} f_\pi p_\mu -i{5\over 18} \sqrt{2} f_\pi \delta^2  
\left ( {1\over 2} x^2 p_\mu  -{1\over 5} x_\mu x\cdot p \right )\ .
\label{expansion}
\end{eqnarray}
Note that higher twists have been neglected as usual in QCD sum rules.
Invoking the isospin symmetry, we directly obtain 
\begin{eqnarray}
\langle 0| {\bar u}(0) \gamma_\mu \gamma_5 u(x) | \pi^0 (p)\rangle
&=& i f_\pi p_\mu -i{5\over 18} f_\pi \delta^2
\left ( {1\over 2} x^2 p_\mu  -{1\over 5} x_\mu x\cdot p \right )\nonumber\ ,\\
\langle 0| {\bar d}(0) \gamma_\mu \gamma_5 d(x) | \pi^0 (p)\rangle
&=& -i f_\pi p_\mu +i{5\over 18} f_\pi \delta^2
\left ( {1\over 2} x^2 p_\mu  -{1\over 5} x_\mu x\cdot p \right )\ .
\label{udexpand}
\end{eqnarray}

\section{Derivation of the pseudotensor matrix elements}
\label{sec:appb}

Here we calculate the matrix element
\begin{eqnarray}
\langle 0| {\bar u}(0) \gamma_5 \sigma_{\alpha \beta} u(x) | \pi^0(p) \rangle
\end{eqnarray}
to leading order in $p_\mu$.
By expanding $u(x)$ in $x_\mu$, we have 
\begin{eqnarray}
\langle 0| {\bar u}(0) \gamma_5 \sigma_{\alpha \beta} u(0) | \pi^0(p) \rangle
+x^\mu \langle 0| {\bar u}(0) \gamma_5 \sigma_{\alpha \beta} 
D_\mu u(0) | \pi^0(p) \rangle + \cdot \cdot \cdot
\ .
\end{eqnarray}
The dots are higher
orders than ${\cal O} (p_\mu)$~\cite{bely}. 
The first term in the expansion is zero simply because it is not 
possible to make antisymmetric combinations  with respect to
$\alpha$ and $\beta$ using the available
vector $p_\mu$ and the metric $g_{\mu\nu}$.
The matrix element in the second term can be written
\begin{eqnarray}
\langle 0| {\bar u}(0) \gamma_5 \sigma_{\alpha \beta} 
D_\mu u(0) | \pi^0(p) \rangle
= i (p_\alpha g_{\beta \mu} - p_\beta g_{\alpha \mu}) A\ .
\end{eqnarray}
No other combinations are
allowed at order ${\cal O} (p_\mu)$.
To get the scalar function $A$, we 
multiply both sides
with $g_{\mu\beta}$ and get, 
\begin{eqnarray}
-i\langle 0| {\bar u}(0) \gamma_5  
D_\alpha u(0) | \pi^0(p) \rangle + {\cal O} (m_q)
= 3ip_\alpha A\ .
\end{eqnarray}
The $m_q$ term is higher chiral order than $p_\mu$ and
can be neglected at the order that we are interested in.
The other term in the LHS is proportional to the first moment of the
twist-3 pion wave function,
\begin{eqnarray}
\langle 0| {\bar u}(0) \gamma_5 D_\alpha u(0) | \pi^0(p) \rangle
={\langle{\bar u} u\rangle \over f_\pi} p_\alpha 
\int_0^1 du~u~\varphi_p(u)
={\langle{\bar u} u\rangle \over 2 f_\pi} p_\alpha \ .
\end{eqnarray}
Note that the zeroth moment of the wave function, that is $\int_0^1 du~\varphi_p(u)=1$, is 
fixed solely by the soft-pion theorem. The first moment
$\int_0^1 du~u~\varphi_p(u)=1/2$ has been used according to Ref.~\cite{bely}.
Hence, 
\begin{equation}
A= -{\langle{\bar u} u\rangle \over 6 f_\pi}\ ,
\end{equation}
which yields the tensor matrix element at the order of ${\cal O}( p_\mu)$,
\begin{eqnarray}
\langle 0|{\bar u}(0) \gamma_5 \sigma_{\alpha \beta} u(x) | \pi^0(p) \rangle
= -i(p_\alpha x_\beta - p_\beta x_\alpha) 
{\langle{\bar u} u\rangle \over 6 f_\pi}\ .
\end{eqnarray}
Due to the isopin symmetry, the $d$-quark element
is given by
\begin{eqnarray}
\langle 0| {\bar d}(0) \gamma_5 \sigma_{\alpha \beta} d(x) | \pi^0(p) \rangle
= i(p_\alpha x_\beta - p_\beta x_\alpha)
{\langle{\bar d} d\rangle \over 6 f_\pi}\ .
\end{eqnarray}
The sign different from the u-quark element can be directly seen
by using the soft-pion theorem.

\section{Derivation of the three-particle pion matrix elements }
\label{sec:appc}

In this appendix, we derive the three-particle pion
matrix element, Eq.~(\ref{three}), contributing  
to our sum rules to leading order in $p_\mu$.
This matrix element can be obtained by inserting a gluonic
tensor from a quark propagator into the quark-antiquark component
with a pion.  Among various possibilities, the one
that survives to leading order in $p_\mu$ can be written,
\begin{eqnarray}
\langle 0| q(x)^\alpha_a i g_s [{\tilde G}^A (0)]^{\sigma \rho} 
{\bar q}(0)^\beta_b | \pi^0(p) \rangle
= t^A_{ab} (\gamma_\theta)^{\alpha \beta} i (A^q_G)^{\theta \sigma\rho}\ .
\label{ths}
\end{eqnarray}
Here ${\tilde G}^A_{\alpha\beta}$ is the dual of the gluon 
strength tensor, 
${\tilde G}^A_{\alpha\beta}={1\over 2} 
\epsilon_{\alpha\beta\sigma\rho} (G^A)^{\sigma\rho}$, and
the color matrix $t^A$ is related to the Gell-Mann matrices via
$t^A=\lambda^A/2$. Other possibilities in combining a gluonic
tensor with the quark-antiquark component are at least the second
order in $p_\mu$~\cite{bely}. 
On multiplying $t^A_{ba} (\gamma^\delta)^{\beta\alpha}$ on both sides,
Eq.~(\ref{ths}) becomes,
\begin{eqnarray}
-\langle 0|{\bar q}(0)\gamma^\delta ig_s {\tilde {\cal G}}^{\sigma\rho} (0)
q(x)| \pi^0(p) \rangle
= 16 i(A_G^q)^{\delta\sigma\rho}\ ,
\label{whatis}
\end{eqnarray}
where ${\tilde {\cal G}}_{\sigma\rho}=t^A {\tilde G}^A_{\sigma\rho}$. 
At order $p_\mu$, the matrix element 
of the LHS contributes at the
local limit ($x_\mu =0$).
Among all possible antisymmetric combinations
with respect to the indices $\sigma$ and $\rho$, the only 
possibility is
\begin{eqnarray}
\langle 0|{\bar q}(0)\gamma^\delta ig_s {\tilde {\cal G}}^{\sigma\rho} (0)
q(0)| \pi^0(p) \rangle = (p^\rho g^{\sigma\delta} - p^\sigma g^{\rho\delta})
A^q_G\ .
\label{th1}
\end{eqnarray}
To determine the scalar function $A^q_G$, we multiply both sides
with $g^{\delta\sigma}$.  Then after some manipulations,
the LHS becomes
\begin{eqnarray}
\langle 0|{\bar q}(0)\gamma_\sigma ig_s {\tilde {\cal G}}^{\sigma\rho} (0)
q(0)| \pi^0(p) \rangle &=&
\langle 0|{\bar q}(0)\gamma^\rho \gamma_5 i D^2 q(0)| \pi^0(p) \rangle 
\nonumber \\
&=&\pm \delta^2 f_\pi p^\rho
\label{th2}
\end{eqnarray}
where the plus sign is
for the $u$-quark and the minus sign is for the $d$-quark.
From Eqs. (\ref{th1}) and (\ref{th2}), we have $A^q_G = \pm \delta^2 f_\pi/3$,
which from Eq.(\ref{whatis}) yields
\begin{eqnarray}
i(A_G^q)^{\delta\sigma\rho} = \mp {1\over 48}
(p^\rho g^{\sigma\delta} - p^\sigma g^{\rho\delta}) \delta^2 f_\pi\ .
\end{eqnarray} 
Therefore, putting into Eq.~(\ref{ths}), we have Eq.~(\ref{three}).

\end{appendix}

\begin{table}
\caption{
The best-fitting values for the parameters $a$ and $b$ 
from the the $i\gamma_5 {\hat p}$
sum rules in the SU(3) symmetric limit.
The Borel window is chosen $0.65 \le M^2 \le 1.24$ GeV$^2$ 
and the continuum threshold we use is $S_0=2.07$ GeV$^2$.
The numbers in the parenthesis are obtained with the
slightly larger continuum threshold threshold $S_0=2.57$
GeV$^2$.
In the fourth column, we present the ratio of each
coupling divided by $g_{\pi N}$, which
is directly related to the parameter $\alpha= F/(F+D)$.
All ratios in the fourth column satisfy the SU(3) relation for the
couplings but the obtained $\alpha$ is highly
sensitive to the input parameters.
The obtained value is  $\alpha=1.236$ when 
$S_0 = 2.07$ GeV$^2$ is used but $\alpha=0.288$ when $S_0 =2.57$ GeV$^2$. 
It can be shown that the value of $\alpha$ 
is also sensitive to the chosen Borel window.
Therefore, it may not be meaningful to determine the $F/D$ ratio from the 
$i\gamma_5 {\hat p}$ sum rule.}

\begin{center}
\begin{tabular}{cccc}
& $a$ (GeV$^6$) & $b$ (GeV$^4$) & coupling$/g_{\pi N}$ \\
\hline\hline
$\pi NN$ & $0.00029 (-0.00158)$ & $0.01055 (0.014)$ & $ 1 $ \\
$\eta NN$ & $0.00067 (-0.00014)$ & $0.00502 (0.00649)$ & $ 2.31 (0.09) $  \\
$\pi \Xi\Xi$ & $0.00043 (0.00067)$ & $-0.00093 (-0.00137)$ & $1.48 (-0.42)$  \\
$\eta \Xi\Xi$ & $-0.00059 (0.00144)$ & $-0.01164 (-0.01535)$ & $-2.03 (-0.91)$  \\
$\pi \Sigma\Sigma$ & $0.00073 (-0.00091)$ & $0.00962 (0.0126)$ & $2.52 (0.58) $  \\
$\eta \Sigma\Sigma$ & $-0.00008 (-0.0013)$ & $0.00663 (0.00886)$ & $-0.28
(0.82)$ 
 \\
\end{tabular}
\end{center}
\label{tab1}

\end{table}

\begin{table}
\caption{The SU(3) symmetric results for 
the $\gamma_5 \sigma_{\mu\nu} q^\mu p^\nu$
sum rules.  The Borel window is chosen  as
$0.65 \le M^2 \le 1.24$ as before.
The numbers in the parenthesis
are when the slightly larger continuum threshold $S_0 =2.57$ 
GeV$^2$ is used.  
In contrast to Table~\ref{tab1},
we have the results rather insensitive to the continuum threshold.
Again, the ratios in the fourth column are shown to satisfy the
SU(3) relations for the couplings exactly.
}

\begin{center}
\begin{tabular}{cccc}
& $a$ (GeV$^6$) & $b$ (GeV$^4$) & coupling$/g_{\pi N}$ \\
\hline\hline
$\pi NN$ & $0.00451 (0.00434)$ & $0.0017 (0.00202)$ & $ 1 $ \\
$\eta NN$ & $0.00196 (0.00206)$ & $-0.0011 (-0.00127)$ & $ 0.43 (0.47) $  \\
$\pi \Xi\Xi$ & $-0.00055 (-0.00039)$ & $-0.00181 (-0.00211)$ & $ -0.12(-0.09)$ \\
$\eta \Xi\Xi$ & $-0.00488 (-0.00479)$ & $-0.00094 (-0.00112)$ & $-1.08(-1.1)$ \\
$\pi \Sigma\Sigma$ & $0.00395 $ & $-0.00009 $ & $0.88 $  \\
$\eta \Sigma\Sigma$ & $0.00292 (0.00273)$ & $0.00204 (0.00239)$ & $0.64 (0.63)$ 
 \\
\end{tabular}
\end{center}
\label{tab2}

\end{table}

\begin{table}
\caption{The best-fitting parameters $a$ and $b$ when
the SU(3) breaking effects are included. 
The 
Borel windows are taken around the resonance masses.
The continuum thresholds are set as shown to
give a qualitative idea how the couplings change
as the SU(3) breaking effects participate .}

\begin{center}
\begin{tabular}{ccccc}
& $a$ (GeV$^6$) & $b$ (GeV$^4$) & Borel window (GeV$^2$)& $S_0$ (GeV$^2$) \\
\hline\hline
$\eta NN$ & $0.00247$ & $-0.00093$ & $0.65 - 1.24$ & $2.07$ \\
$\pi \Xi\Xi$ & $-0.00284$ & $-0.00264$ & $1.5 - 1.9$ & $3.$  \\
$\eta \Xi\Xi$ & $-0.01194$ & $-0.00099$ & $1.5 - 1.9$ & $3.$  \\
$\pi \Sigma\Sigma$ & $0.00504$ & $-0.00017$ & $1.2 - 1.6$ & -  \\
$ \eta \Sigma \Sigma$ & $0.00473$ & $0.00189$ & $1.2 - 1.6$ & $3.$
 \\
\end{tabular}
\end{center}
\label{tab3}

\end{table}

\begin{table}
\caption{ Meson-baryon diagonal couplings in the SU(3) limit
and beyond the SU(3) limit are presented. For the SU(3) breaking
case, we have 
used the scaling for the strength $\lambda^2_{B_j}\propto m_{B_j}^6$.
The $\pi NN$ coupling
in the first line is the empirical value. We emphasize that,
due to the limitations mentioned in the text,
the values in the third column should be regarded as just qualitative
guide for the SU(3) breaking in the couplings within
our approach. In the 4th and 5th columns, the couplings from
other works are shown for comparison.}

\begin{center}
\begin{tabular}{ccccc}
& SU(3)  & Broken SU(3) & Nijmegen~\cite{rijken1} & BH~\cite{bh}\\
\hline\hline
$g_{\pi N}$ & $13.4$ & $13.4$ & - & - \\
$g_{\eta N}$ & $5.76$ & $7.34$ & $6.37$ & - \\
$g_{\pi \Xi}$ & $-1.61$ & $-1.13$ & $-5.3$ & $-6.05$ \\
$g_{\eta \Xi}$ & $-14.47$ & $-4.73$ & $-11.09$ & - \\
$g_{\pi \Sigma}$ & $11.79$ & $3.66$ &$11.75$ & $9.24$ \\
$ g_{\eta \Sigma} $ & $8.58$ & $3.43 $ & $15.53$ & -
 \\
\end{tabular}
\end{center}
\label{tab4}

\end{table}

\begin{figure}
\caption{ The Borel curves for the $\gamma_5 \sigma_{\mu\nu} q^\mu p^\nu$
sum rules in the SU(3) limit. 
The dashed lines show the sensitivity to the continuum threshold.}
\label{fig3}
\end{figure}

\setlength{\textwidth}{6.1in}   
\setlength{\textheight}{9.in}  
\begin{figure}
\centerline{%
\vbox to 2.4in{\vss
   \hbox to 3.3in{\includegraphics{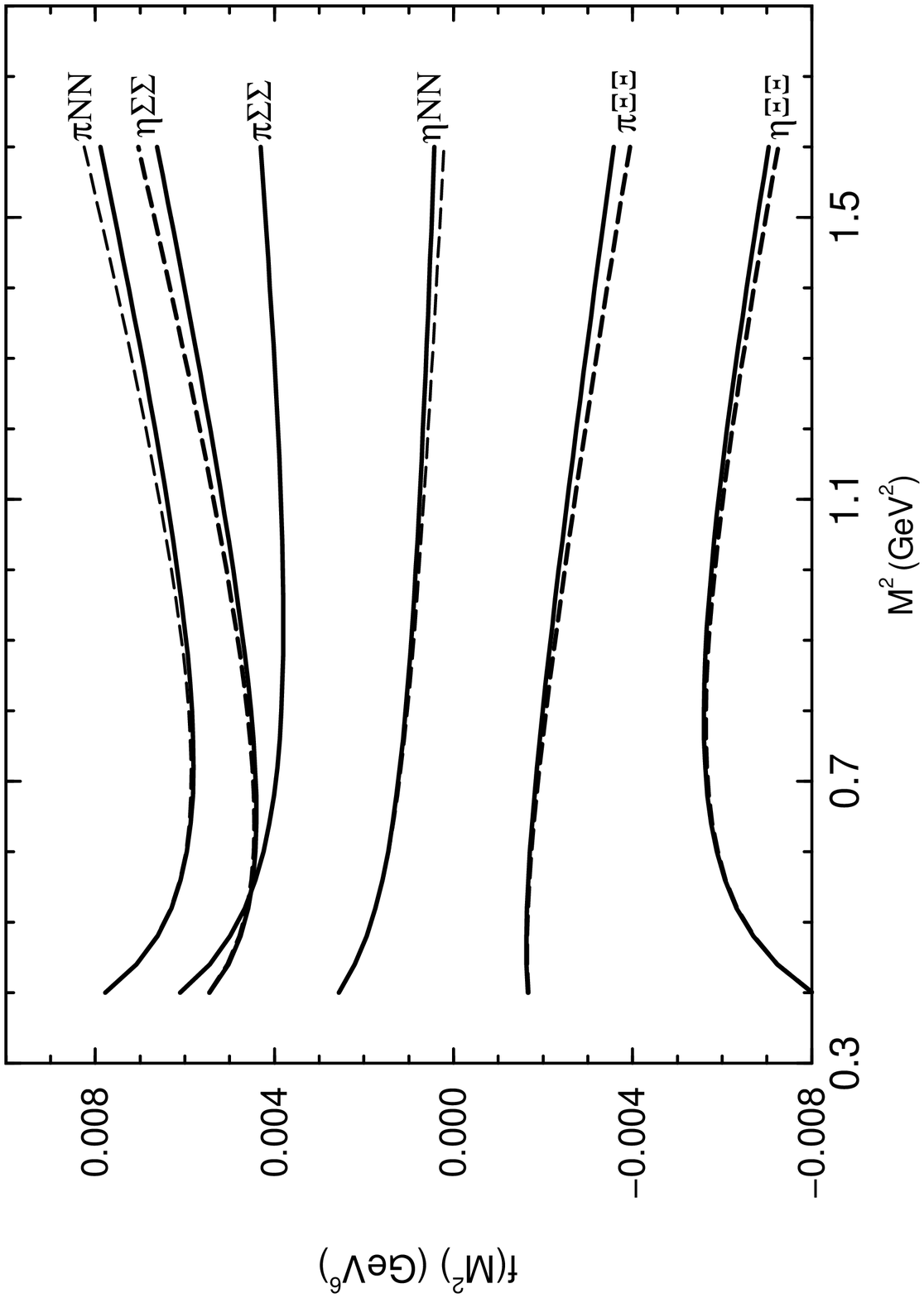}\hss}}
}
\eject
\end{figure}

\begin{figure}
\caption{ The Borel curves of the 
$\gamma_5 \sigma_{\mu\nu} q^\mu p^\nu$
sum rules for $\pi \Sigma \Sigma$, $\eta \Sigma \Sigma$,
$\pi \Xi \Xi$ and $\eta \Xi \Xi$ when the SU(3) breaking
effects are included.
The continuum threshold $S_0 = 3$ GeV$^2$ has been
used for the plots but the sensitivity of the curves to this choice 
is small.
}
\label{fig4}

\setlength{\textwidth}{6.1in}   
\setlength{\textheight}{9.in}  
\centerline{%
\vbox to 2.4in{\vss
   \hbox to 3.3in{\includegraphics{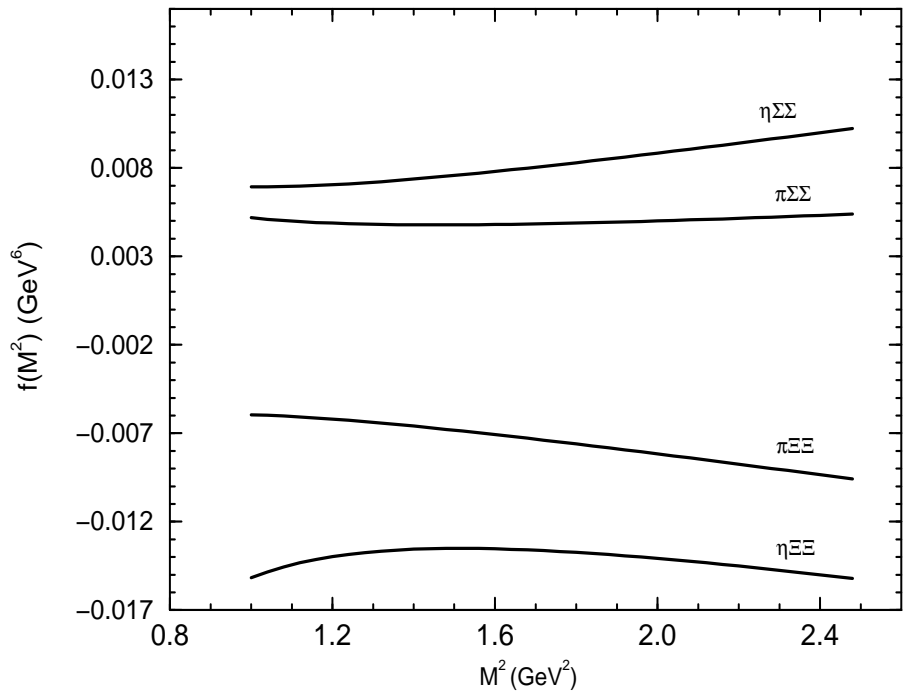}\hss}}
}
\vspace{130pt}
\end{figure}

\end{document}